\documentclass[aps,prd,preprint,floatfix]{revtex4}
\usepackage{epsfig}
\usepackage{latexsym}

\begin{document}

\preprint{\rightline{ANL-HEP-PR-04-64}}
 
\title{The finite temperature transition for 2-flavour lattice QCD at finite 
isospin density.}
 
\author{J.~B.~Kogut}
\address{Dept. of Physics, University of Illinois, 1110 West Green Street,
Urbana, IL 61801-3080, USA}
\author{D.~K.~Sinclair}
\address{HEP Division, Argonne National Laboratory, 9700 South Cass Avenue,
Argonne, IL 60439, USA}
 
\begin{abstract}
We simulate 2-flavour lattice QCD at finite isospin chemical potential $\mu_I$,
for temperatures close to the finite temperature transition from hadronic
matter to a quark-gluon plasma. The $\mu_I$ dependence of the transition
coupling is observed and used to estimate the decrease in the transition
temperature with increasing $\mu_I$. These simulations are performed on an
$8^3 \times 4$ lattice at 3 different quark masses. Our estimate of the 
magnitude of the fluctuations of the phase of the fermion determinant at small
quark-number chemical potential $\mu$, suggest that the position of the small
$\mu$ and small $\mu_I$ transitions should be the same for $\mu_I=2\mu$, and
we argue that the nature of these transitions should be the same. For all
$\mu_I < m_\pi$ the smoothness of these transitions and the values of the
Binder cumulant $B_4$, indicate that these transitions are mere crossovers, and
show no sign of the expected critical endpoint. For $\mu_I > m_\pi$ and a small
isospin ($I_3$) breaking term $\lambda$, we do find evidence of a critical
endpoint which would indicate that, for $\lambda=0$, there is a tricritical
point on the phase boundary where the pion condensate evaporates, where this
phase transition changes from second to first order.
\end{abstract}

\maketitle

\section{Introduction}

QCD at finite baryon-/quark-number density describes nuclear matter. Beyond
nuclei it describes the physics of neutron stars and has the potential to
predict such exotic objects as quark stars. Hot hadronic matter at low 
baryon-number density was present in the early universe. Relativistic heavy-ion
collisions at RHIC and CERN produce hot nuclear matter.

QCD at a finite chemical potential $\mu$ for quark-number, has a complex fermion
determinant, which makes the naive application of standard lattice simulation
methods, which are based on importance sampling, difficult if not impossible.
To circumvent these problems people have introduced various schemes which are
applicable to high temperatures and small $\mu$. These include various
reweighting techniques \cite{Fodor:2001au,Fodor:2001pe,Fodor:2004nz}, 
and methods which expand physical observables as
power series in $\mu$ \cite{Allton:2002zi,Karsch:2003va,deForcrand:2002ci,
deForcrand:2003hx,D'Elia:2002gd,Gavai:2003mf,Gavai:2003nn} or related 
parameters \cite{Azcoiti}.

Another approach is to study theories which are expected to possess some of
the properties of QCD at finite $\mu$, but have real positive fermion
determinants, making them amenable to standard simulation methods. One
such theory is QCD at finite chemical potential $\mu_I$ for isospin ($I_3$).
This theory has been studied both by effective (chiral) Lagrangian techniques
\cite{Son:2000xc,Son:2000by},
as well as by direct lattice simulations \cite{Kogut:2002zg}. 
At zero temperature these studies
indicate that this theory undergoes a second-order phase transition with
mean-field critical exponents at $\mu_I=m_\pi$, to a state characterized by a
charged pion condensate which breaks $I_3$ spontaneously.

We report here a study of 2-flavour lattice QCD at finite $\mu_I$ and finite
temperature ($T$), in the neighbourhood of the finite temperature transition
from hadronic matter to a quark-gluon plasma. Since we work at finite quark
mass to make the pion massive and thus to move pion condensation to finite
$\mu_I$, the finite temperature transitions form a line of crossovers
emanating from the $\mu_I=0$ transition, for small $\mu_I$. We calculate the
position of this crossover as a function of $\mu_I$ on an $8^3 \times 4$
lattice for 3 different quark masses ($m=0.05,0.1,0.2$), from the peaks of
the susceptibilities of the various observables, using Ferrenberg-Swendsen
reweighting to interpolate between the $\beta$ values used in our simulations.
For $\mu_I < m_\pi$, we set the symmetry breaking parameter $\lambda=0$. From
estimations of the fluctuations of the phase of the fermion determinant for
small quark-number chemical potential $\mu$ we shall argue that there is an
appreciable range of $\mu$ over which these fluctuations are small enough that
the position of the crossover at finite $\mu$ will be the same as that at
finite $\mu_I$ with $\mu_I=2\mu$. We find good agreement with the $\mu$
dependence of this transition observed by de Forcrand and Philipsen
\cite{deForcrand:2002ci}. This agreement between the $\mu$ and $\mu_I$
dependence of the transition $\beta=6/g^2$ and hence temperature was noted by
the Bielefeld-Swansea group \cite{Allton:2002zi}. We also find that the
transition for each of our 3 masses appears to remain a crossover with no sign
of a critical endpoint for all $\mu_I < m_\pi$. Preliminary results from these
simulations have been presented at conferences
\cite{Kogut:2002se,Kogut:2003cd,Sinclair:2003rm}.

We have also studied the finite temperature transition for $\mu_I > m_\pi$.
Here, for symmetry breaking parameter $\lambda=0$, the pion condensate
evaporates at the finite temperature transition, which is thus a true phase
transition. However, since here the fermion propagator becomes singular (at
least in the infinite volume limit) for temperatures below this transition,
because of the Goldstone boson associated with the spontaneous breaking of
$I_3$, we are forced to work at finite (small) $\lambda$, where the transition
again becomes a crossover. Here we shall present evidence for a critical 
endpoint beyond which the transition becomes first order. Because $\lambda$ is
small, we shall argue that this first order behaviour persists to $\lambda=0$.
At $\lambda=0$, the finite temperature crossover is replaced by a second order
transition, the first order transition remains first order and the critical
endpoint becomes a tricritical point. Although most of our simulations were
performed on $8^3 \times 4$ lattices, we performed some simulations on $16^3
\times 4$ lattices close to the critical endpoint.

In section 2, we introduce lattice QCD at finite $\mu_I$. In section 3 we
define the fourth order Binder cumulants which we use to study the nature of
the transitions. Section 4 describes our simulations and results for small
$\mu_I$ ($\mu_I < m_\pi$). The large $\mu_I$ simulations and results are 
presented in section 5. Section 6 contains discussions and conclusions.

\section{Lattice QCD at finite $\mu_I$}

The staggered quark action for lattice QCD at finite chemical potential $\mu_I$ 
for isospin ($I_3$) is
\begin{equation}
S_f=\sum_{sites} \left[\bar{\chi}[D\!\!\!\!/(\frac{1}{2}\tau_3\mu_I)+m]\chi
                   + i\lambda\epsilon\bar{\chi}\tau_2\chi\right],
\end{equation}
where $D\!\!\!\!/(\frac{1}{2}\tau_3\mu_I)$ is the standard staggered quark
transcription of $D\!\!\!\!/$ with the links in the $+t$ direction multiplied 
by $\exp(\frac{1}{2}\tau_3\mu_I)$ and those in the $-t$ direction multiplied by
$\exp(-\frac{1}{2}\tau_3\mu_I)$. The term proportional to $\lambda$ is an
explicit $I_3=\frac{1}{2}\tau_3$ symmetry breaking term. This term serves two
purposes. Firstly, such a term is necessary if one is to see evidence for 
spontaneous $I_3$ breaking on a finite lattice. Secondly, it prevents the
Dirac operator from becoming singular, as we see below. $\tau_1$, $\tau_2$ and
$\tau_3$ are the isospin matrices so that this Dirac operator is a 
$2 \times 2$ matrix in isospin space. The determinant
\begin{equation}
\det[D\!\!\!\!/(\frac{1}{2}\tau_3\mu_I) + m + i\lambda\epsilon\tau_2]
               =\det[{\cal A}^\dagger{\cal A}+\lambda^2],
\label{eqn:det}
\end{equation}
where
\begin{equation}
{\cal A} \equiv D\!\!\!\!/(\frac{1}{2}\mu_I)+m ,
\end{equation}
is a $1 \times 1$ matrix in isospin space, which means that we only need use
a single flavour-component fermion field in our simulations. This determinant
is real and positive allowing us to use standard hybrid molecular-dynamics 
simulations, with noisy fermions to allow us to tune the number of flavours
from 8 down to 2.

We note that, for $\lambda=0$, the determinant of equation~\ref{eqn:det} is
just the magnitude of the determinant for 8-flavour lattice QCD with
quark-number chemical potential
\begin{equation} 
\mu=\frac{1}{2}\mu_I .
\end{equation}

Observables for this theory include the chiral condensate,
\begin{equation}
\langle\bar{\psi}\psi\rangle \Leftrightarrow \langle\bar{\chi}\chi\rangle,
\end{equation}
the charged pion condensate
\begin{equation}
i\langle\bar{\psi}\gamma_5\tau_2\psi\rangle  \Leftrightarrow
i\langle\bar{\chi}\epsilon\tau_2\chi\rangle
\end{equation}
and the isospin density
\begin{equation}
j_0^3 = \frac{1}{V}\left\langle{\partial S_f \over \partial\mu_I}\right\rangle.
\end{equation}
We will also be interested in the Wilson Line (Polyakov Loop), and the
plaquette observable
\begin{equation}
{\rm PLAQUETTE}= S_\Box = 1-\frac{1}{3}{\rm Re}{\rm Tr}U_\Box .
\end{equation}
 
\section{Fourth-order Binder cumulants}

If $X$ is an observable, its 4-th order Binder cumulant is defined by
\begin{equation}
B_4 = {\langle(X - \langle X \rangle)^4\rangle \over
       \langle(X - \langle X \rangle)^2\rangle^2}
\end{equation}
which approaches a universal value at a critical point \cite{binder}. It has
been pointed out if one chooses $X$ to be an eigenvector of the critical
scaling Hamiltonian, $B_4$ will be as close as is possible to its infinite
volume limit on finite volumes \cite{Karsch:2001nf}. If one plots $B_4$ as a
function of those parameters which parameterize the departure from the
critical point, the curves obtained for different lattice sizes will intersect
at the critical point if $X$ is indeed an eigenvector. For other choices of
$X$ the intersections of such curves will only tend to this unique value in
the infinite volume limit. The value of the cumulant at this point of
intersection will be that characteristic of the universality class of this
critical point and the nature of the observable.

For transitions other than critical points, the Binder cumulant only attains
its characteristic value in the infinite volume limit. For a crossover, the
infinite volume value for the Binder cumulant for the order parameter
is $B_4=3$. For a first order transition, this Binder cumulant is $B_4=1$. The 
critical endpoint we are seeking is expected to be in the universality class of
the 3-dimensional Ising model for which $B_4=1.604(1)$. For a mean field 
critical point $B_4=\Gamma(5/4)\Gamma(1/4)/\Gamma(3/4)^2=2.1884...$ for a
1-component order parameter \cite{lipowski}, or $B_4=\pi/2=1.570796...$ for a 2
component order parameter. At a 3-dimensional tricritical point for a
1-component order parameter $B_4=2$ \cite{lipowski}, and
$B_4=\Gamma(1/3)/\Gamma(2/3)^2=1.460998...$ for a 2-component order parameter.

\section{Simulations and results for $\mu_I < m_\pi$}

We have simulated 2-flavour QCD on $8^3 \times 4$ lattices in the neighbourhood
of the finite temperature transition from hadronic matter to a quark-gluon
plasma, for small values of the isospin ($I_3$) chemical potential $\mu_I$.
Here small $\mu_I$ means $\mu_I < m_\pi$ so that, even at zero temperature,
the system is in the normal phase, i.e. in the phase where there is no 
$I_3$-breaking charged pion condensate. We set $\lambda=0$, since a finite
$\lambda$ is only needed when there is a possibility of spontaneous $I_3$
breaking. Simulations were performed at 3 different quark masses 
$m=0.05,0.1,0.2$.

At the lowest quark mass $m=0.05$ we performed simulations over a range of
$\mu_I$ values $0 \leq \mu_I \leq 0.55$, where the highest $\mu_I$ value is
only just below the critical $\mu_I$ above which a pion condensate forms at low
temperature. The larger quark masses were chosen to allow an even larger
range of $\mu_I$s for the normal phase. For $m=0.1$ we simulated over the 
range $0 \leq \mu_I \leq 0.7$, while for $m=0.2$ we covered the range
$0 \leq \mu_I \leq 1$. At each of the selected $\mu_I$ values we performed
simulations over a range of $\beta=6/g^2$ values spanning the finite 
temperature transition. We used a range of updating increments (in 
molecular-dynamics `time') for these simulations. These ranged from $dt=0.1$
for $m=0.1, 0.2$ and small $\mu_I$ down to $dt=0.01$ for $m=0.05$ and 
$\mu_I=0.55$ close to the transition. For each quark mass we performed
simulations as long as $20,000$ molecular-dynamics time units (trajectories)
at individual values of $(\beta,\mu_I)$ close to a transition.

\begin{figure}[htb]
\epsfxsize=6in
\centerline{\epsffile{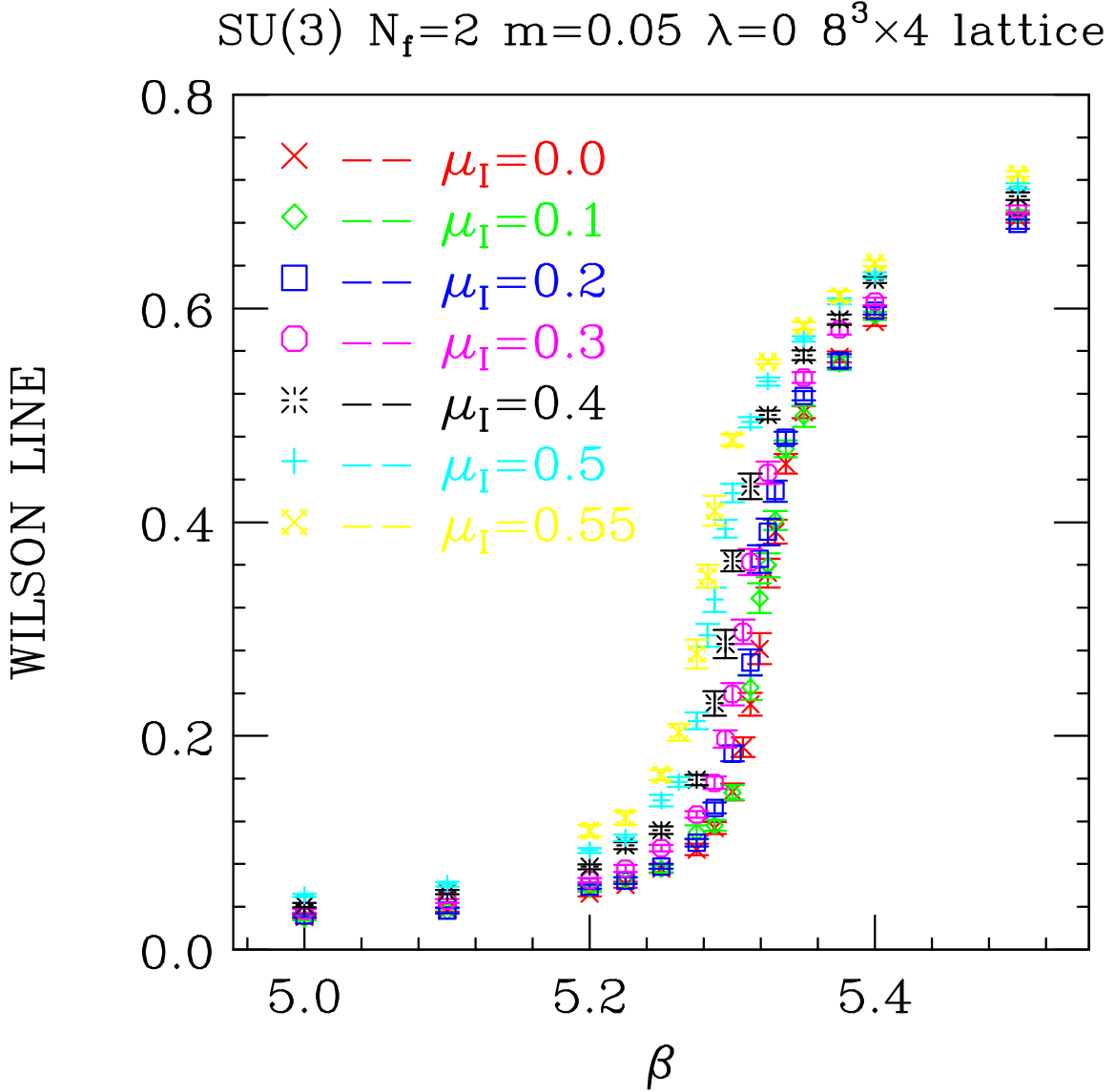}}
\caption{Wilson Line as a function of $\beta$ for various $\mu_I < m_\pi$ and
$m=0.05$.}\label{fig:wilson0.05}
\end{figure}

\begin{figure}[htb]                                                          
\epsfxsize=6in                                  
\centerline{\epsffile{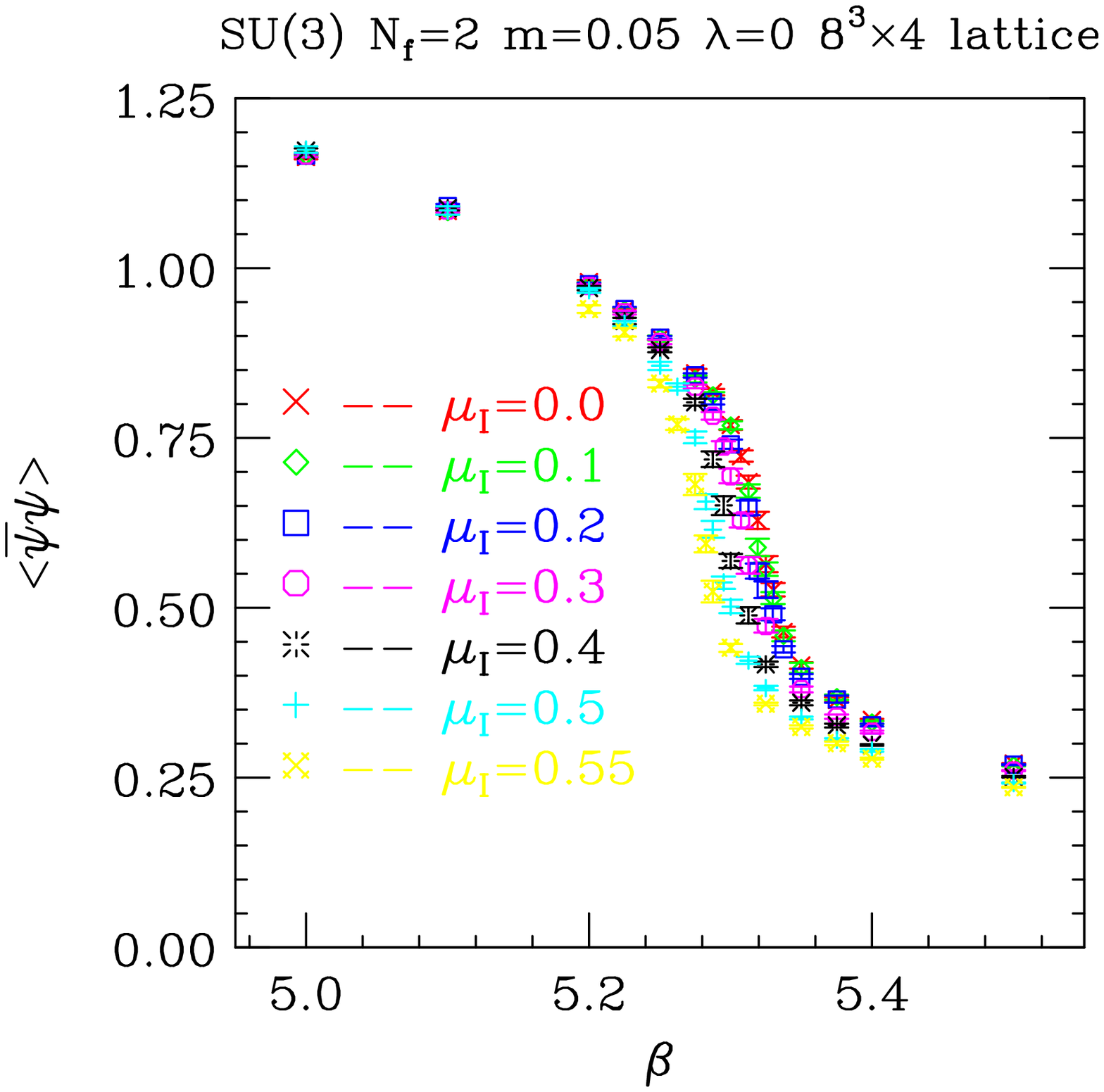}}                   
\caption{$\langle\bar{\psi}\psi\rangle$ as a function of $\beta$ for various 
$\mu_I < m_\pi$ and $m=0.05$.}\label{fig:pbp0.05}     
\end{figure}

\begin{figure}[htb]
\epsfxsize=6in
\centerline{\epsffile{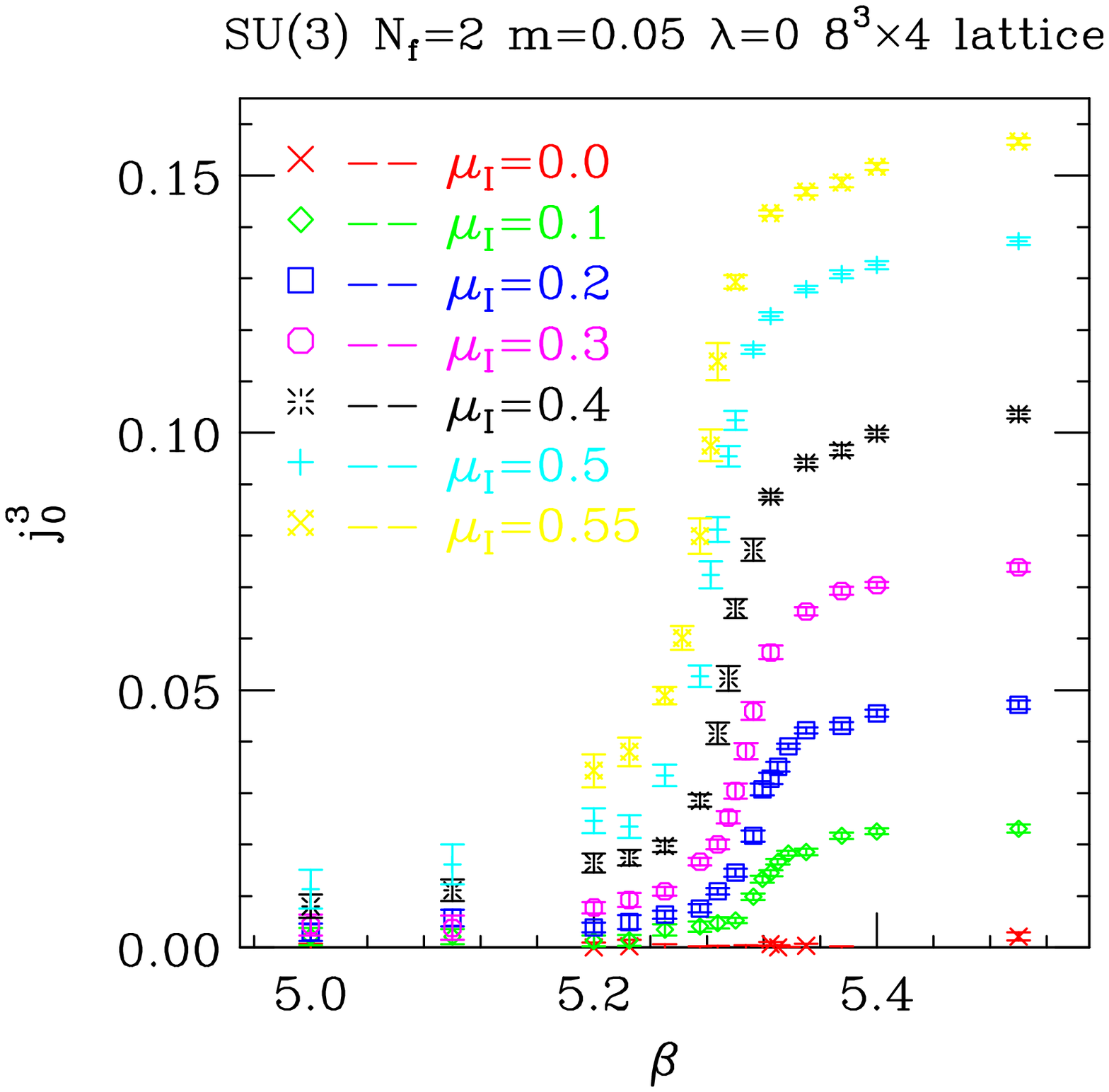}}   
\caption{$j_0^3$ as a function of $\beta$ for various $\mu_I < m_\pi$ and 
$m=0.05$.}\label{fig:j0_0.05}
\end{figure}

At each value of $m$, $\mu_I$ and $\beta$ we measured the average plaquette,
the Wilson Line (Polyakov Loop), the chiral condensate and the isospin density
for each trajectory. For the fermionic quantities, where we calculate
stochastic estimates, we used 5 noise vectors for each trajectory, which
enabled us to make unbiased estimates of the susceptibilities and Binder
cumulants. Figure~\ref{fig:wilson0.05} shows the Wilson Line as a function of
$\beta$ for each $\mu_I$ at quark mass $m=0.05$.
Note that there is a rapid crossover marking the transition. In addition we
see that the position of the crossover moves towards smaller $\beta$ and hence
lower temperature as $\mu_I$ is increased. However, we notice that the 
crossover $\beta$, $\beta_c$ varies only slowly with $\mu_I$. The corresponding
values of the chiral condensate, $\langle\bar{\psi}\psi\rangle$ are given in
figure~\ref{fig:pbp0.05}. Again we see a rapid crossover close to the position
of that for the Wilson Line.
Figure~\ref{fig:j0_0.05} shows the behaviour of the isospin density $j_0^3$
for the same mass and $\mu_I$s. Here we see the finite temperature transition
again. We note that the value of the isospin density in the quark-gluon plasma
(high $\beta$) increases with increasing $\mu_I$. At each $\mu_I$ it appears
to level off at large $\beta$. Note that this is not the lattice artifact of
saturation; $j_0^3$ in this domain is far below its saturation value of $3$.
The rise in $j_0^3$ occurs because increasing $\mu_I$ raises the Fermi surface.
These observables for $m=0.1$ and $m=0.2$ behave very similarly to those for
$m=0.05$, except that the crossovers occur at larger $\beta$ values as mass is
increased.

The transitions we have observed in each of these masses and chemical potentials
appear to be smooth crossovers rather than actual phase transitions, as is
expected to be the case for $\mu_I=0$ (we will present further evidence for
this later in this section). The position of the transition is thus defined
as the $\beta$ value which maximizes a chosen susceptibility. (Such definitions
and Ferrenberg-Swendsen reweighting are used by other groups, as are the 
Binder cumulant methods used below \cite{Allton:2002zi,Karsch:2003va,
deForcrand:2002ci,deForcrand:2003hx,Karsch:2001nf}.)This is a 
reasonable definition only if the positions of the maxima of the 
susceptibilities for the various observables are close, at least in the 
infinite volume limit. The susceptibility for a chosen observable ${\cal O}$
is defined as
\begin{equation}
\chi_{\cal O} = V \langle {\cal O}^2 - \langle {\cal O} \rangle^2 \rangle,
\end{equation}
Where V is the space-time volume of the lattice. Note that this is the correct
definition only for a local observable ${\cal O}$. We have also used this
definition for the Wilson Loop which is only local in 3-space. Thus what we
call $\chi_{Wilson}$ is strictly $N_t\chi_{Wilson}=\chi_{Wilson}/T$. 

\begin{figure}[htb]
\epsfxsize=6in
\centerline{\epsffile{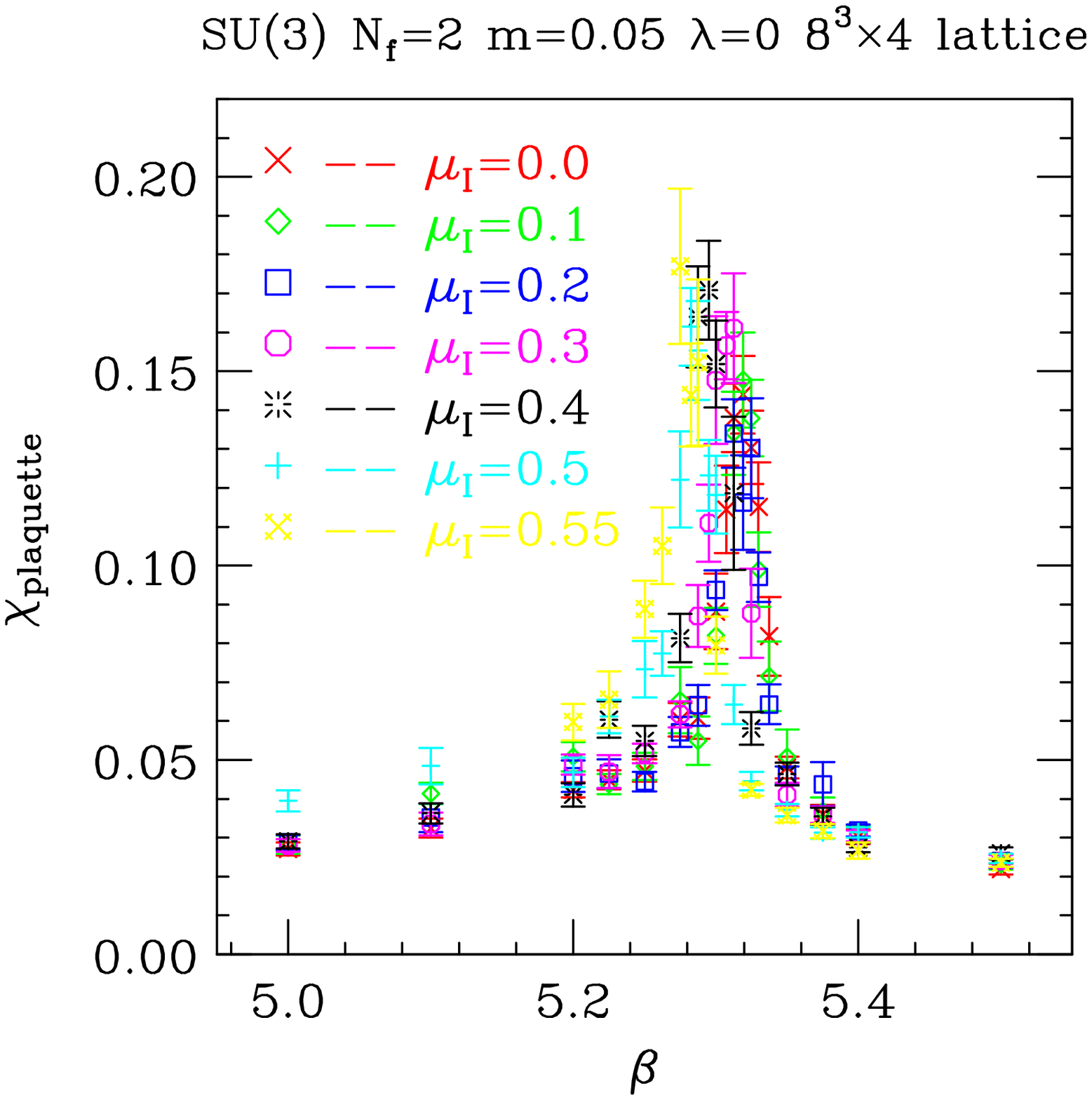}}
\caption{Plaquette susceptibilities as functions of $\beta$ for $\mu_I < m_\pi$,
for $m=0.05$.}\label{fig:chiplaq}
\end{figure}

\begin{figure}[htb]
\epsfxsize=6in
\centerline{\epsffile{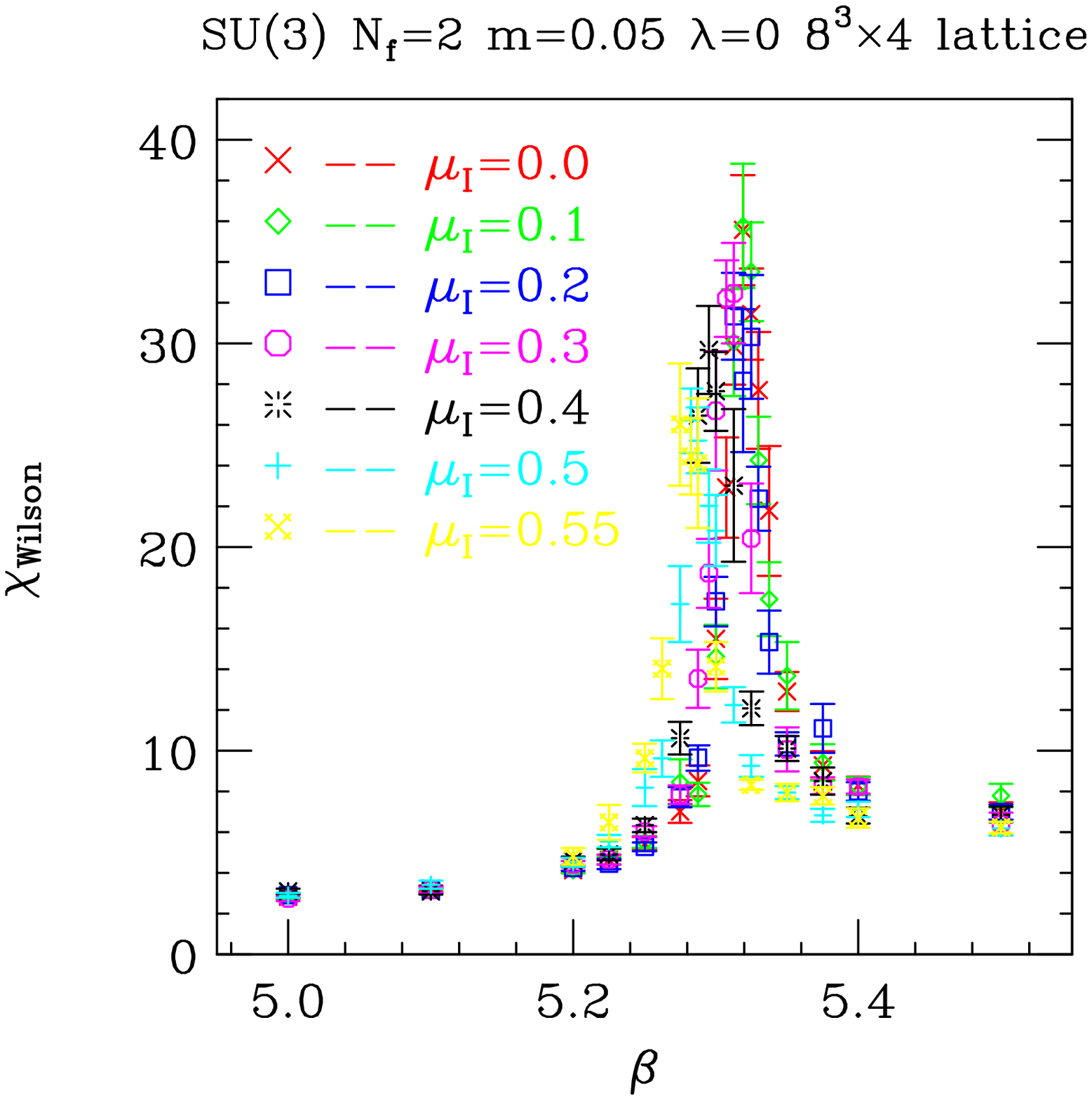}}
\caption{Wilson line susceptibilities as functions of $\beta$ for 
$\mu_I < m_\pi$, for $m=0.05$.}\label{fig:chiwilson}                        
\end{figure}

\begin{figure}[htb]                                                       
\epsfxsize=6in                                                           
\centerline{\epsffile{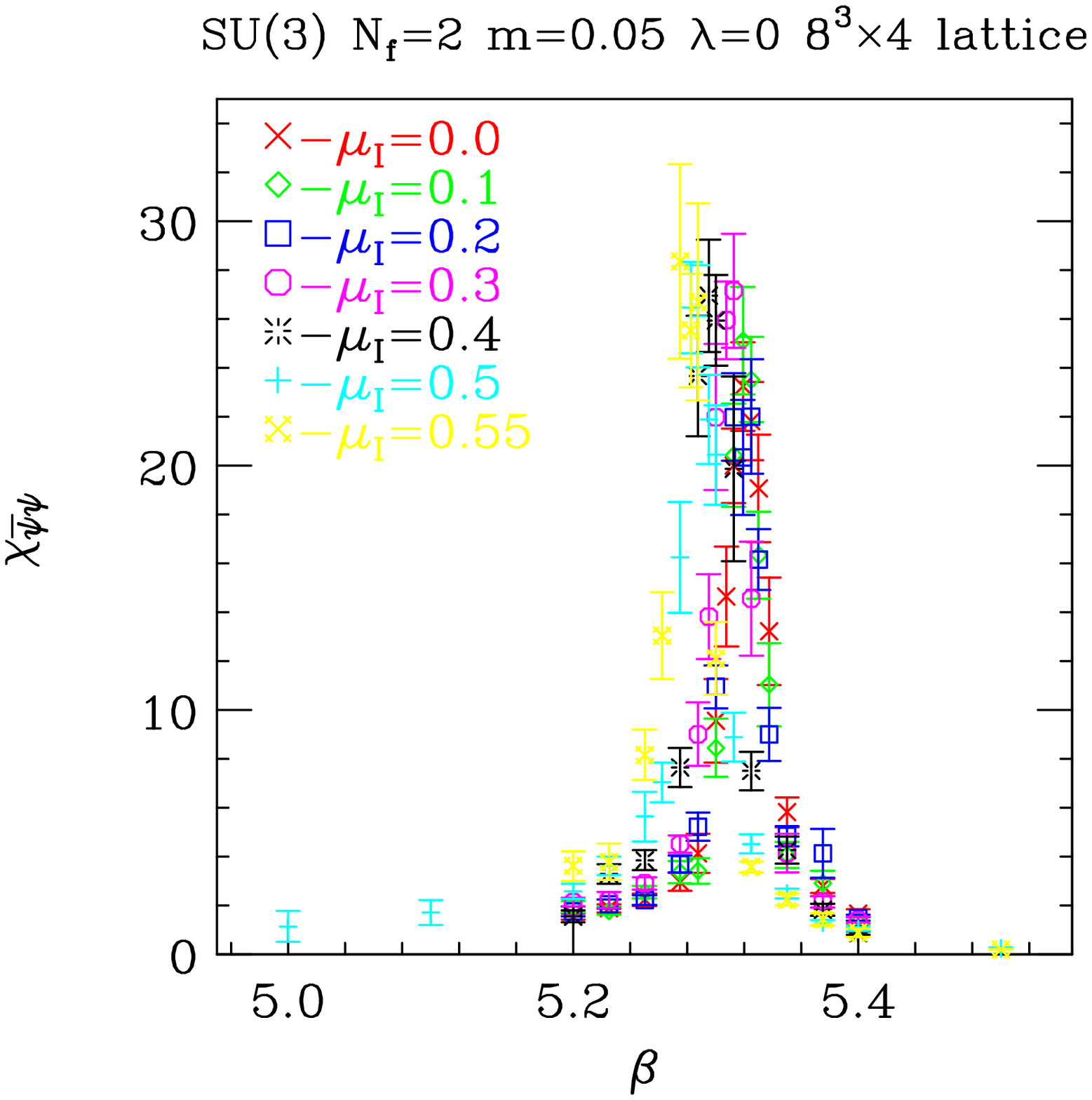}}
\caption{$\bar{\psi}\psi$ susceptibilities as functions of $\beta$ for 
$\mu_I < m_\pi$, for $m=0.05$.}
\label{fig:chipbp}                                               
\end{figure}

\begin{figure}[htb]                                                         
\epsfxsize=6in                                                                
\centerline{\epsffile{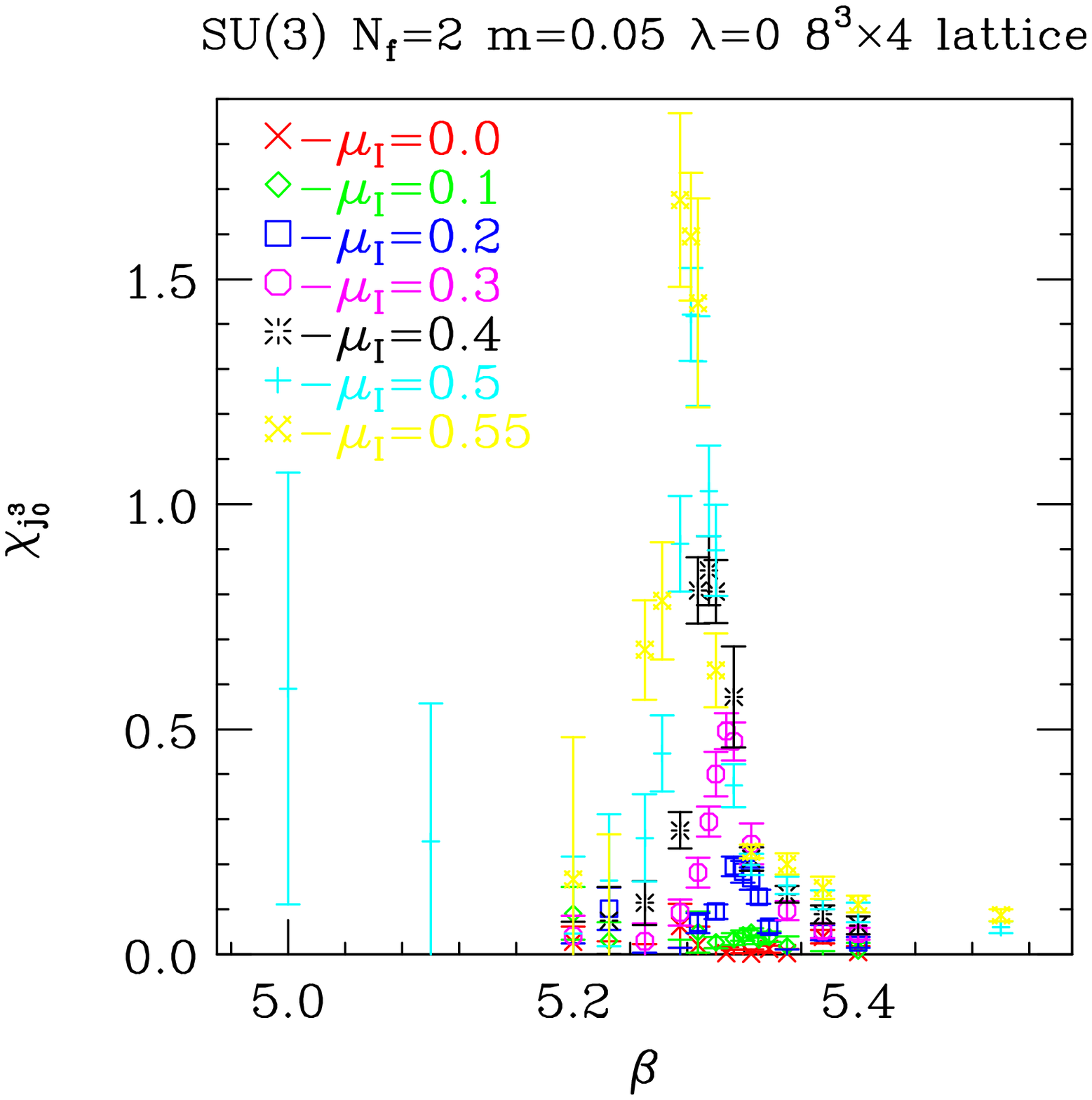}}            
\caption{Isospin density susceptibilities as functions of $\beta$ for 
$\mu_I < m_\pi$, for $m=0.05$.}\label{fig:chij0}                       
\end{figure}

Figure~\ref{fig:chiplaq} shows the plaquette susceptibilities for $m=0.05$
at those $\beta$s at which we performed our simulations for each $\mu_I$.
These susceptibilities are clearly strongly peaked, and the peaks move to
lower $\beta$s as $\mu_I$ is increased. Figure~\ref{fig:chiwilson} gives the
corresponding susceptibilities for the Wilson/Polyakov Line.
Again these susceptibilities are strongly peaked and the peak moves to lower
$\beta$ as $\mu_I$ is increased. The main difference is that the peaks in these
susceptibilities decrease slightly in height as $\mu_I$ is increased, whereas
those for the plaquette susceptibilities increase with increasing $\mu_I$.
Figure~\ref{fig:chipbp} shows the susceptibilities for the chiral condensate,
also for $m=0.05$. By using all 5 stochastic estimators of 
$\langle\bar{\psi}\psi\rangle$ and removing the noise-diagonal contribution 
we obtain an unbiased estimate of this susceptibility
These susceptibilities are the most strongly peaked of all the susceptibilities
and the height of the peaks increases with increasing $\mu_I$. Finally we show
the susceptibilities for the isospin densities at $m=0.05$ in 
figure~\ref{fig:chij0}. Again we used the 5 noise vectors to obtain an unbiased
estimator.
The rapid increase in the height of these peaks with increasing $\mu_I$ 
corresponds to the increase in $j_0^3$ seen in figure~\ref{fig:j0_0.05}.

In order to pinpoint the susceptibility peaks more precisely, we use the
distribution of observables and plaquette actions measured during our runs
and use Ferrenberg-Swendsen reweighting \cite{Ferrenberg:yz} to estimate the
susceptibilities at $\beta$ values close to those at which we have performed
simulations. If ${\cal O}$ is an observable for which ${\cal O}_i$, $i=1,...n$
are the measured values (lattice averaged), and $S_{\Box\:i}$ are the
corresponding plaquette actions, at $\beta=\beta_0$, then
\begin{equation}
\langle{\cal O}\rangle(\beta) = {\sum_i \exp(-6\,V (\beta-\beta_0) S_{\Box\:i}) 
\: {\cal O}_i             \over  \sum_i \exp(-6\,V (\beta-\beta_0) S_{\Box\:i})}
\end{equation}
for $\beta$ close enough to $\beta_0$ that the distributions of $S_\Box$ values
at $\beta$ and $\beta_0$ have significant overlap. Applying this formula to
estimate both $\langle{\cal O}\rangle$ and $\langle{\cal O}^2\rangle$ yields
the desired susceptibility $\chi_{\cal O}$. Jackknife methods are used to 
determine the errors in both the susceptibilities and the positions of their
peaks. It turns out that for our simulations at each $m$ and $\mu_I$, we have
$3-5$ $\beta$ values close enough to the peak $\beta_c$ to be used to determine
$\beta_c$. After checking that the estimates of $\beta_c$ from each of these
points are consistent, we obtain our final estimate as a $\chi^2$ weighted
average of these.

\begin{figure}[htb]
\epsfxsize=6in
\centerline{\epsffile{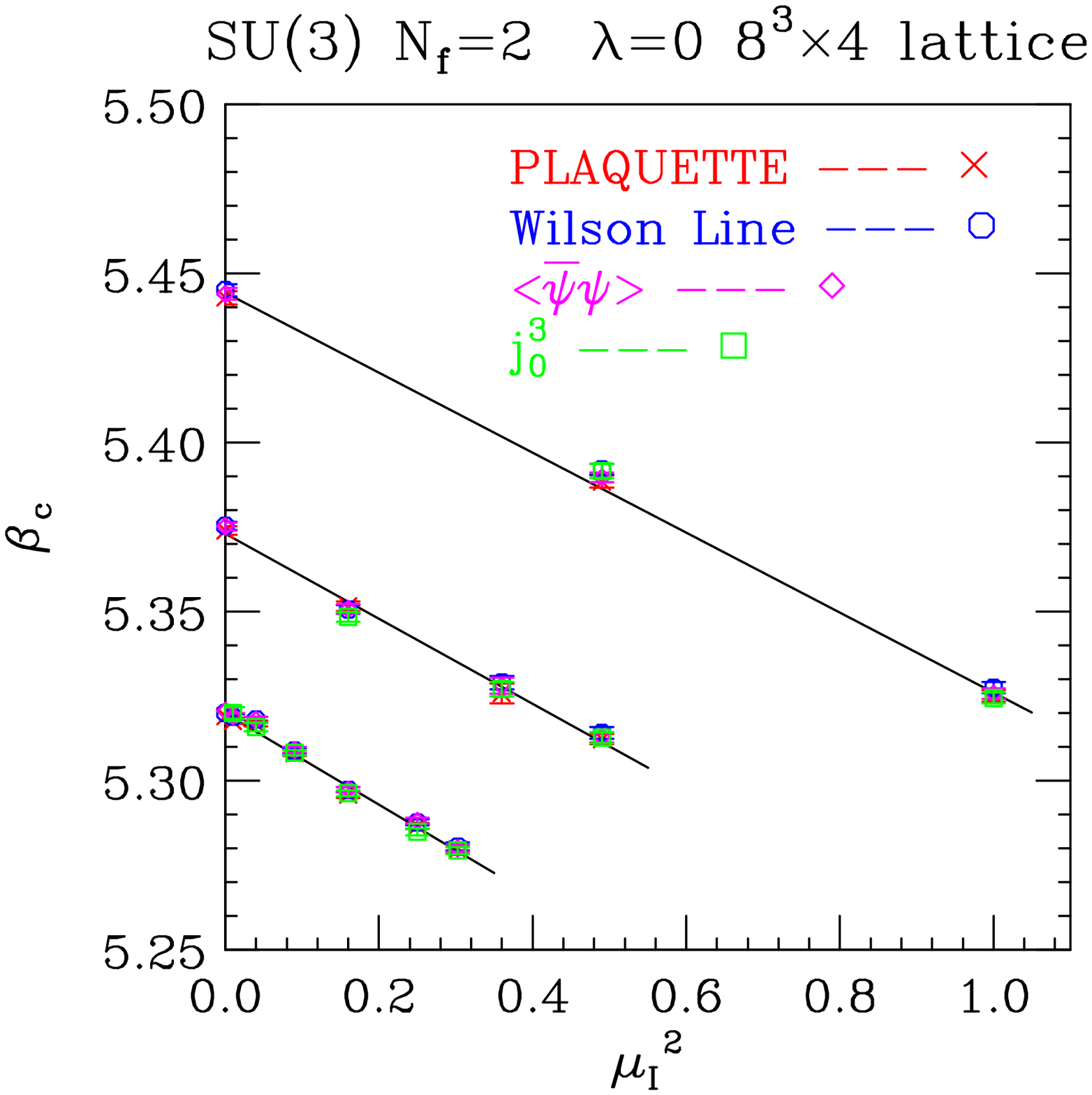}}
\caption{$\beta_c$ as functions of $\mu_I^2$ together with straight line fits
for each mass. The bottom set of points and line are for $m=0.05$. The middle
set of points and line are for $m=0.1$. The top set of points and line are for 
$m=0.2$.}\label{fig:beta}
\end{figure}

In figure~\ref{fig:beta} we plot our $\beta_c$ estimates from each of the 4
susceptibilities as functions of $\mu_I^2$ for all 3 masses.
There is clearly good agreement between the $\beta_c$ values obtained from the 
different susceptibilities. Arguments as to why this should be so have been
presented in references \cite{Hatta:2003ga,Fukushima:2002mp,Mocsy:2003tr,
Mocsy:2003qw}. Since for a fixed mass
\begin{equation}
\beta_c(\mu_I) = \beta_c(-\mu_I)
\end{equation}
$\beta_c$ is a function of $\mu_I^2$. For $|\mu_I|$ small enough it is also an
analytic function of $\mu_I$. Thus for small $\mu_I$, 
\begin{equation}                                                              
\beta_c(\mu_I) = a + b \mu_I^2 + c \mu_I^4 + ...
\end{equation} 
We have therefore fit $\beta_c$ to the form $\beta_c(\mu_I) = a + b \mu_I^2$,
for each mass. Choosing to fit the plaquette susceptibilities, we get
\begin{eqnarray}
\beta_c &=& 5.3198(9)\;\: - 0.134(6) \mu_I^2 \;\;\;\; for \; m=0.05 \nonumber\\
\beta_c &=& 5.3731(11)    - 0.126(5) \mu_I^2 \;\;\;\; for \; m=0.1  \nonumber\\
\beta_c &=& 5.4443(27)    - 0.118(4) \mu_I^2 \;\;\;\; for \; m=0.2. \nonumber\\
\label{eqn:beta_c}
\end{eqnarray}
These lines are plotted in figure~\ref{fig:beta}. Since $\chi^2/dof$ is $3.2$
for $m=0.05$, $1.1$ for $m=0.1$ and $2.4$ for $m=0.2$, these are not excellent
fits. However, including the quartic term, does not improve the $m=0.05$ fit,
and adding even more terms would be a meaningless exercise.

It was observed by the Bielefeld-Swansea group that at small $\mu$ and $\mu_I$,
the dependence of $\beta_c$ on $\mu$ and on $\mu_I$ was identical within their
errors \cite{Allton:2002zi}. 
In our notation this would mean that
\begin{equation}
\beta_c(\mu) = \beta_c(\mu_I=2\mu).
\end{equation}
This means that one can determine the position of the finite temperature 
transition for small $\mu$ by simulating with the magnitude of the fermion
determinant and ignoring the phase. Let us give an intuitive argument why this
might be the case. If $\theta$ is the phase of the fermion determinant, and
${\cal O}$ is a gauge-field observable (such as the plaquette or Wilson line)
then
\begin{equation}
\langle{\cal O}\rangle_\mu = {\langle\exp(i\,\theta){\cal O}\rangle_{\mu_I=2\mu}
\over                         \langle\exp(i\,\theta)\rangle_{\mu_I=2\mu}},
\label{eqn:<O>}
\end{equation}
where for ${\cal O}$ real we can replace $\exp(i\,\theta)$ with $\cos(\theta)$.
For the lattice size used by the Bielefeld-Swansea group, there was a range of
$\mu$ over which this denominator was not too small and varied smoothly with
$\mu$ and slowly with $\beta$, as $\beta$ swept across the $\beta_c$.
Here the denominator cannot be responsible for the transition. When the
fluctuations in $\theta$ are so well behaved, it is reasonable to treat
$\exp(i\,\theta){\cal O}$ as another observable. Since the position of the
transition appears to be nearly independent of the chosen observable, this
would suggest that the position of the transition would be the same for this
new observable. If so, the smooth behaviour of the denominator would imply
that the position of the transition at finite $\mu$ should be the same as that
for the transition at finite $\mu_I$ for $\mu_I=2\mu$. It is also not
unreasonable to assume that the nature of the 2 transitions might be the same.
Such observations are not new (see for example \cite{Takaishi:2004qa}).

If the relation between $\beta_c(\mu)$ and $\beta_c(\mu_I)$ holds with the
standard staggered action (the Bielefeld-Swansea group used the $p$-$4$ action)
we can compare our formulae for $\beta_c(\mu_I)$ (equation~\ref{eqn:beta_c})
with that obtained by de Forcrand and Philipsen \cite{deForcrand:2002ci}
\begin{equation}
\beta_c = 5.2865(18) - 0.149(10) \mu_I^2 \;\;\;\; for \; m=0.025
\label{eqn:dfp}
\end{equation}
where we have made the substitution $\mu=\mu_I/2$ in their equation. This would
appear to be consistent with our equations, taking into account the difference
in mass. To examine whether this agreement is quantitative, we fit 
equations~\ref{eqn:beta_c},\ref{eqn:dfp} to the expected scaling form
\begin{eqnarray}
\beta_c(m,\mu_I) &=& \beta_c(m,0) + a(m) \mu_I^2                \nonumber  \\
\beta_c(m,0)     &=& \beta_c(0,0) + b m^\frac{1}{\beta_m\delta} \nonumber  \\
a(m)             &=& a(0)         + c m^\frac{1}{\beta_m\delta}. 
\end{eqnarray}
Such scaling fits have been considered by \cite{Karsch:1994hm,Karsch:2000kv}
at zero chemical potentials. For the expected continuum $O(4)$ scaling
$1/\beta_m\delta \approx 0.55$, while for the lattice $O(2)$ scaling
$1/\beta_m\delta \approx 0.59$. [Note that such scaling is only derivable for
the case of finite $\mu$, where, in the chiral limit, the line of crossovers
becomes a line of second order transitions in the same universality class as
the $\mu=0$($\mu_I=0$) transition. We are using the assumed relationship
between finite $\mu$ and finite $\mu_I$ to extend it to finite $\mu_I$.] The
fit of all 4 equations to $O(4)$ scaling gives $\beta_c(0,0)=5.210(3)$,
$b=0.57(1)$, with a $\chi^2/dof=1.6$, and $a(0)=0.152(6)$, $c=0.85(19)$ with a
$\chi^2/dof=0.5$. The fit to $O(2)$ scaling gives $\beta_c(0,0)=5.219(3)$,
$b=0.59(2)$ with a $\chi^2/dof=2.4$ and $a(0)=0.151(6)$, $c=0.087(20)$ with a
$\chi^2/dof=0.5$. Considering the quality of the fits in
equations~\ref{eqn:beta_c}, we consider either of these scaling fits to be
good enough to support our claim that we are consistent with de Forcrand and
Philipsen, and that the combined measurements are consistent with the expected
scaling with quark mass $m$. Note that our value of $\beta_c(0,0)$ is less
than that obtained in \cite{Karsch:1994hm,Karsch:2000kv}. We should not expect
good agreement with the later paper, since it uses a larger lattice and finite
size effects are non-negligible on an $8^3 \times 4$ lattice. The fit in the
earlier work was over the mass range $0.02 \leq m \leq 0.075$, while ours was
over the range $0.025 \leq m \leq 0.2$. Considering the rapid variation of the
scaling form at small $m$, the difference between our result, $5.210(3)$ and
their's, $5.222(3)$ is perhaps not surprising.

What remains to be checked is that the phase ($\theta$) of the fermion
determinant is well behaved. Since calculating the fermion determinant is very
expensive, we use the series expansion for $\theta$ given in
\cite{Allton:2002zi}. In our normalization,
\begin{equation}
\theta=\frac{1}{4} \mu_I V {\rm Im}(j_0) + {\cal O}(\mu_I^3)
\end{equation}
where $j_0$ is the number density normalized to 4 flavours (1 staggered fermion
field). We use our 5 stochastic estimators/configuration of $j_0$ to obtain an
unbiased estimator of $\langle\theta^2\rangle$ through order $\mu_I^2$. (We
also made an unbiased estimator of $\langle j_0^4 \rangle$ which had a poor
enough signal/noise ratio that we did not even try to estimate 
$\langle\theta^4\rangle$ or the higher order contributions to 
$\langle\theta^2\rangle$.) Our results for a range of $\beta$ values which
span the $\mu_I=0$ transition for each quark mass are given in 
table~\ref{tab:phase}.
\begin{table}[htb]
\begin{tabular}{|c|c|c|c|}
\hline
$m$      &   $\beta$   & $\langle[{\rm Im}(j_0)]^2\rangle$ &
                         $\langle\theta^2\rangle/\mu_I^2$             \\
\hline
0.05     &   5.3000    & $2.1(8) \times 10^{-5}$ &    5.5(2.1)        \\
0.05     &   5.3075    & $2.6(8) \times 10^{-5}$ &    6.7(1.7)        \\
0.05     &   5.3125    & $1.0(5) \times 10^{-5}$ &    2.6(1.2)        \\
0.05     &   5.3190    & $2.1(5) \times 10^{-5}$ &    5.5(1.2)        \\
0.05     &   5.3250    & $1.5(4) \times 10^{-5}$ &    4.0(1.2)        \\
0.05     &   5.3300    & $1.0(5) \times 10^{-5}$ &    2.7(1.2)        \\
0.05     &   5.3375    & $1.0(3) \times 10^{-5}$ &    2.7(0.9)        \\
\hline                                                                
0.10     &   5.3500    & $1.6(4) \times 10^{-5}$ &    4.1(1.2)        \\
0.10     &   5.3625    & $1.8(3) \times 10^{-5}$ &    4.7(0.8)        \\
0.10     &   5.3750    & $1.3(3) \times 10^{-5}$ &    3.3(0.7)        \\
0.10     &   5.3875    & $0.6(2) \times 10^{-5}$ &    1.5(0.6)        \\
0.10     &   5.4000    & $0.2(7) \times 10^{-5}$ &    0.6(1.8)        \\
\hline                                                                
0.20     &   5.4250    & $1.7(2) \times 10^{-5}$ &    4.4(0.6)        \\
0.20     &   5.4375    & $1.2(2) \times 10^{-5}$ &    3.2(0.5)        \\
0.20     &   5.4500    & $1.0(2) \times 10^{-5}$ &    2.5(0.4)        \\
0.20     &   5.4625    & $0.7(1) \times 10^{-5}$ &    1.9(0.4)        \\
0.20     &   5.4750    & $0.7(2) \times 10^{-5}$ &    1.9(0.5)        \\
\hline
\end{tabular}
\caption{Fluctuations in the phase of the fermion determinant.}
\label{tab:phase}
\end{table}
A reasonable measure of how ``well-behaved'' this phase is, is 
$\langle\cos(\theta)\rangle$. When this quantity is close to $1$, 
the oscillations in phase are small, and it is reasonable to produce
ensembles with the magnitude of the determinant and to include the
phase in the measurement. When this expectation value falls towards zero
$\theta$ is almost uniformly distributed over the interval $(-\pi,+\pi]$,
and the contributions of configurations generated using the magnitude of
the determinant can easily cancel, as they would in the denominator of 
equation \ref{eqn:<O>} for this case. How small $\langle\cos(\theta)\rangle$
can get before generating ensembles without the phase becomes invalid is
a matter of ``experimentation'', but one might expect that 
$\langle\cos(\theta)\rangle > 0.5$ would be a reasonable range over which
we could use this method. To the order in $\mu_I$ to which we work we
must take $\cos(\theta) \approx 1 - \frac{1}{2}\theta^2$. Applying our
criterion to the measurements of table~\ref{tab:phase}, we see that, even
in the worst case we should be able to trust the relationship between
measurements at finite $\mu_I$ and finite $\mu$, out to $\mu_I^2 \approx 0.15$,
i.e. out to $\mu_I \approx 0.4$. Thus it is not unreasonable to assume that
this relation will be a reasonable approximation for most of the region
$\mu_I < m_\pi$.

\begin{figure}[htb]
\epsfxsize=4in
\centerline{\epsffile{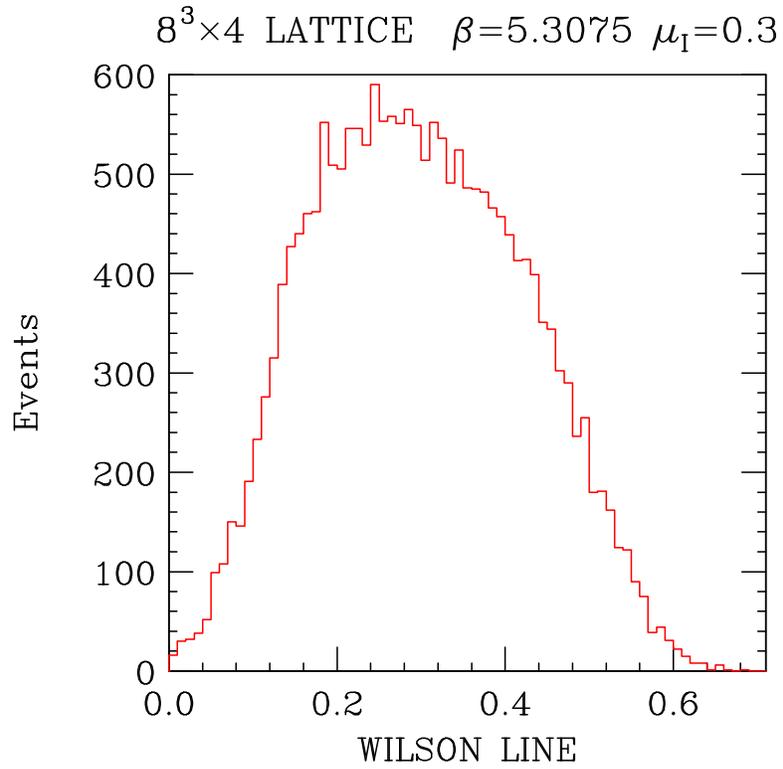}}
\vspace{0.15in}
\centerline{\epsffile{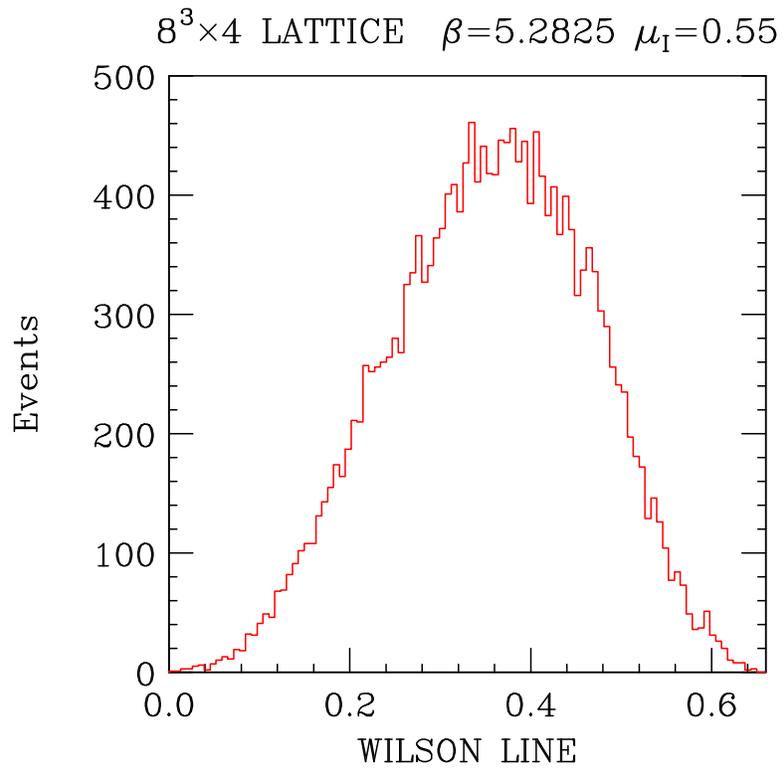}}
\caption{Histograms of distribution of Wilson Line values for $m=0.05$:
a) For $\mu_I=0.3$, $\beta=5.3075$; b) For $\mu_I=0.55$, $\beta=5.2825$.
}\label{fig:hist0}
\end{figure}

Let us now examine the nature of these transitions more closely. We have 
observed that the transitions appear smooth in all measured observables. This
suggests that they are merely rapid crossovers. Histogramming those observables
which could show discontinuities if there were a first order transition, shows
a single broad peak for all masses considered for all $\mu_I < m_\pi$, which
suggests a crossover (or possibly a second order transition) but not a first
order transition. We note that on such small lattices, one can observe a
double peak structure, even where the transition is a crossover or second
order transition. However, it is rare that a first order transition would not
show a double peak, unless it were very weak. In figure~\ref{fig:hist0} we show
histograms of the Wilson Line (Polyakov Loop) for $m=0.05$ at an intermediate
value of $\mu_I$ (0.3) and one close to $m_\pi$ (0.55). These both show a
single broad peak as advertised, and are typical. [We chose to show the
Wilson Line rather than the chiral condensate, since use of stochastic
estimators (even after averaging over all 5 estimates for each configuration)
could possibly obscure a double peak.]

\begin{figure}[htb]
\epsfxsize=6in
\centerline{\epsffile{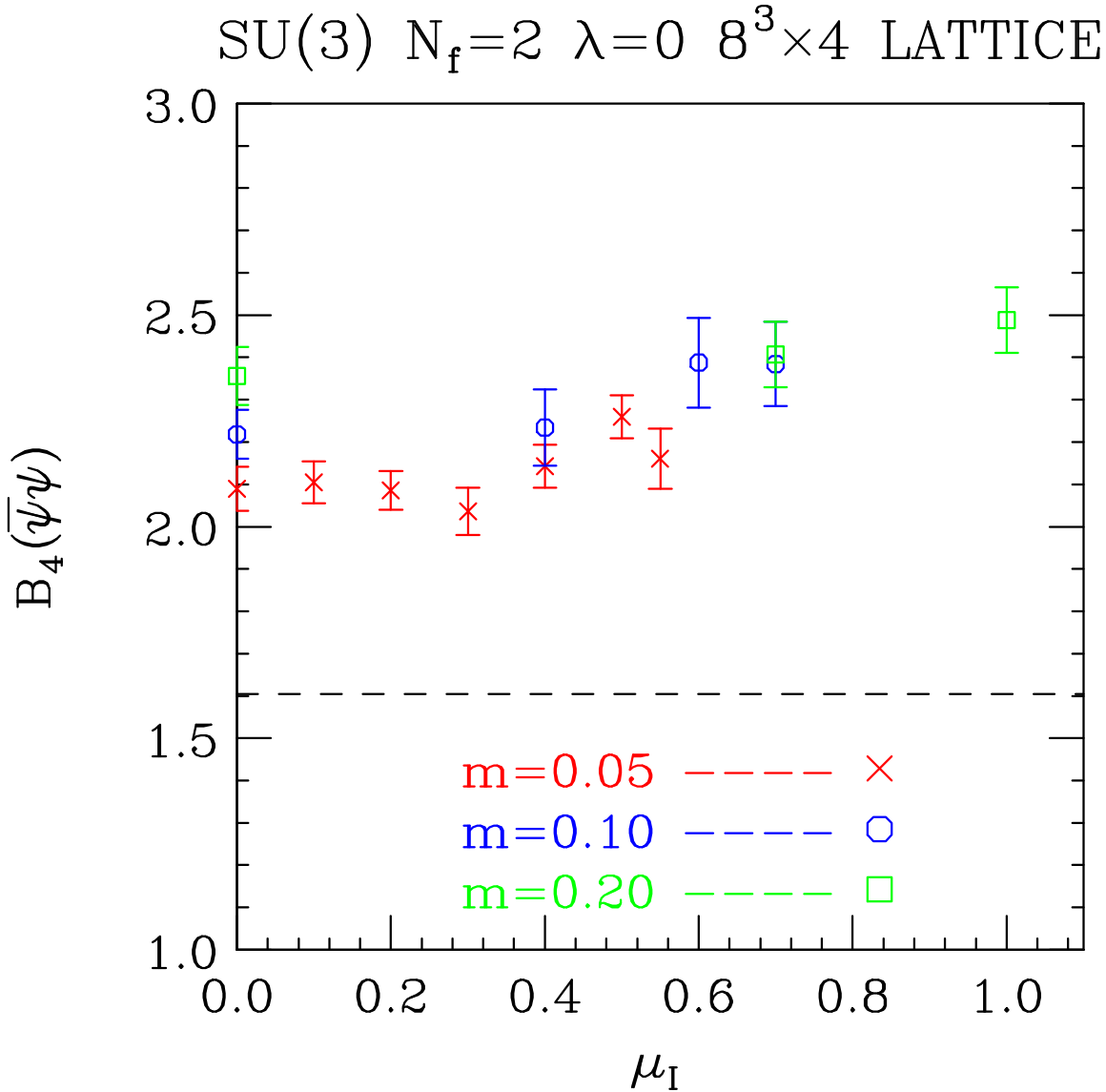}}
\caption{Fourth order Binder cumulants ($B_4$) for 
$\langle\bar{\psi}\psi\rangle$ as a function of $\mu_I$. The dashed line is
at $B_4=1.604$, the value for the 3-dimensional Ising model.}\label{fig:binder0}
\end{figure}

Finally, we have calculated the fourth order Binder cumulants $B_4$ for the
chiral condensate at the transition for each $m$ and $\mu_I$. Having 5 noisy
estimators per configuration, we were able to generate an unbiased estimator
for $B_4$. We again use Ferrenberg-Swendsen reweighting to interpolate between
those $\beta$ values at which we ran our simulations. We determined the
position of the transition for each $m$ and $\mu_I$ as that $\beta$ 
which minimized $B_4$. This method of determining the position of the
transition gave $\beta_c$ values in excellent agreement with those obtained
from the maxima of the corresponding susceptibilities. We plot these $B_4$
values in
figure~\ref{fig:binder0}.

If there were a critical endpoint, which is expected to be in the Ising
universality class, we would expect $B_4$ to pass through its Ising value,
$B_4=1.604$, at this endpoint. For the crossover region, $B_4$ should lie
above the Ising value, approaching $3$ in the limit of large lattices. In the
first order domain (if it existed) $B_4$ should lie below the Ising value,
approaching $1$ in the large lattice limit. We have plotted the Ising value as
a dashed line in figure~\ref{fig:binder0}. Clearly $B_4$ lies well above
$1.604$ and shows no sign of approaching this value. Hence the evidence from
Binder cumulants supports our conclusion that the transition is a crossover
for all $\mu_I < m_\pi$, over a range of quark masses.

\section{Simulations and results for $\mu_I > m_\pi$}

In the region where $\mu_I > m_\pi$,  for $\lambda=0$, the charged pion
condensate evaporates at the finite temperature transition and $I_3$ symmetry,
which is broken spontaneously at low temperature, is restored. Hence this
finite temperature transition is a true phase transition. However, since in
this case the low temperature phase has a true Goldstone mode, this would render
the Dirac operator singular (at least in the large volume limit). Hence we must
use a non-zero $\lambda$, which we keep small so that we can infer information
about the $\lambda \rightarrow 0$ limit. For $\lambda \ne 0$, the phase
transition is no longer required, and our earlier work indicates that for
$\mu_I$ just above $m_\pi$, the transition is softened to a crossover. We can
now search for a critical endpoint in the high $\mu_I$ ($\mu_I > m_\pi$)
regime. Again the critical endpoint would be expected to lie in the 
universality class of the 3-dimensional Ising model. For $\mu_I$ above this
endpoint the transition would be first order. Unfortunately, in this domain,
we cannot argue that the finite $\mu_I$ and the finite $\mu$ transitions are
related.

\begin{figure}[htb]
\epsfxsize=6in
\centerline{\epsffile{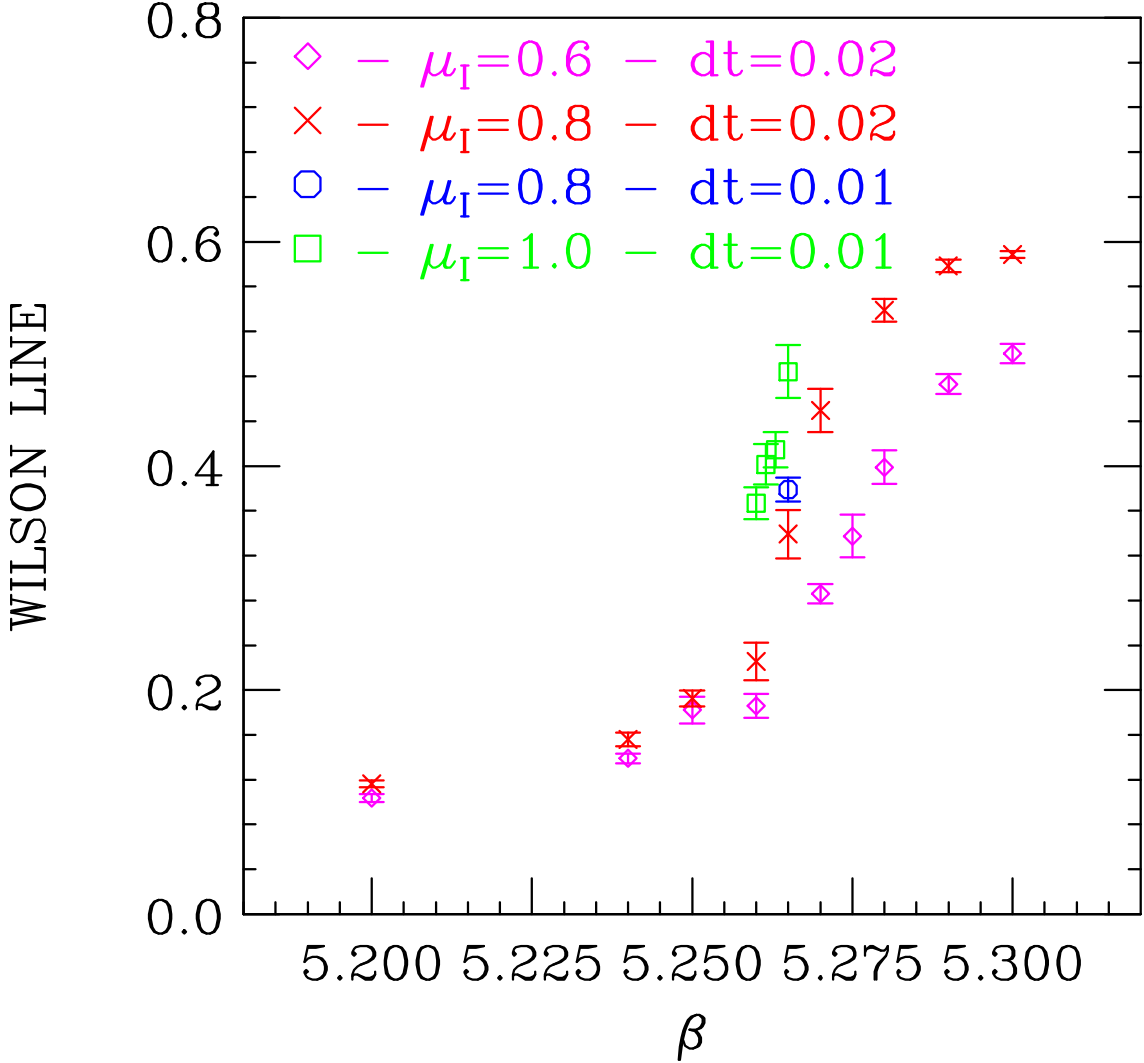}}
\caption{Wilson Line (Polyakov Loop) for large $\mu_I$ on an $8^3 \times 4$ 
lattice with $m=0.05$ and $\lambda=0.005$}\label{fig:wilson_hi}
\end{figure}

\begin{figure}[htb]                                                            
\epsfclipon
\epsfxsize=3.5in                                                      
\hspace{-0.75in} \epsffile{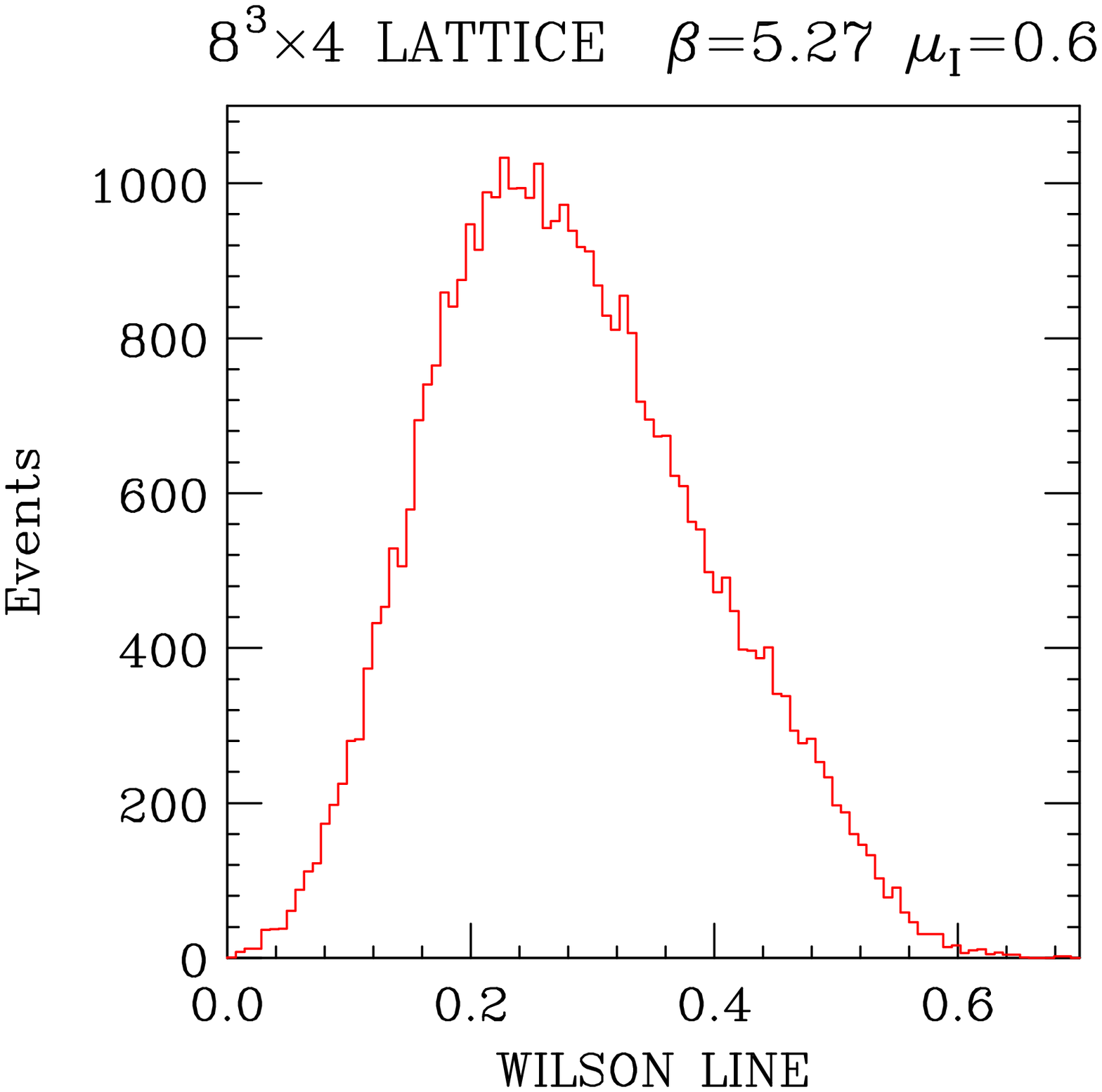}
\epsfxsize=3.5in
\epsffile{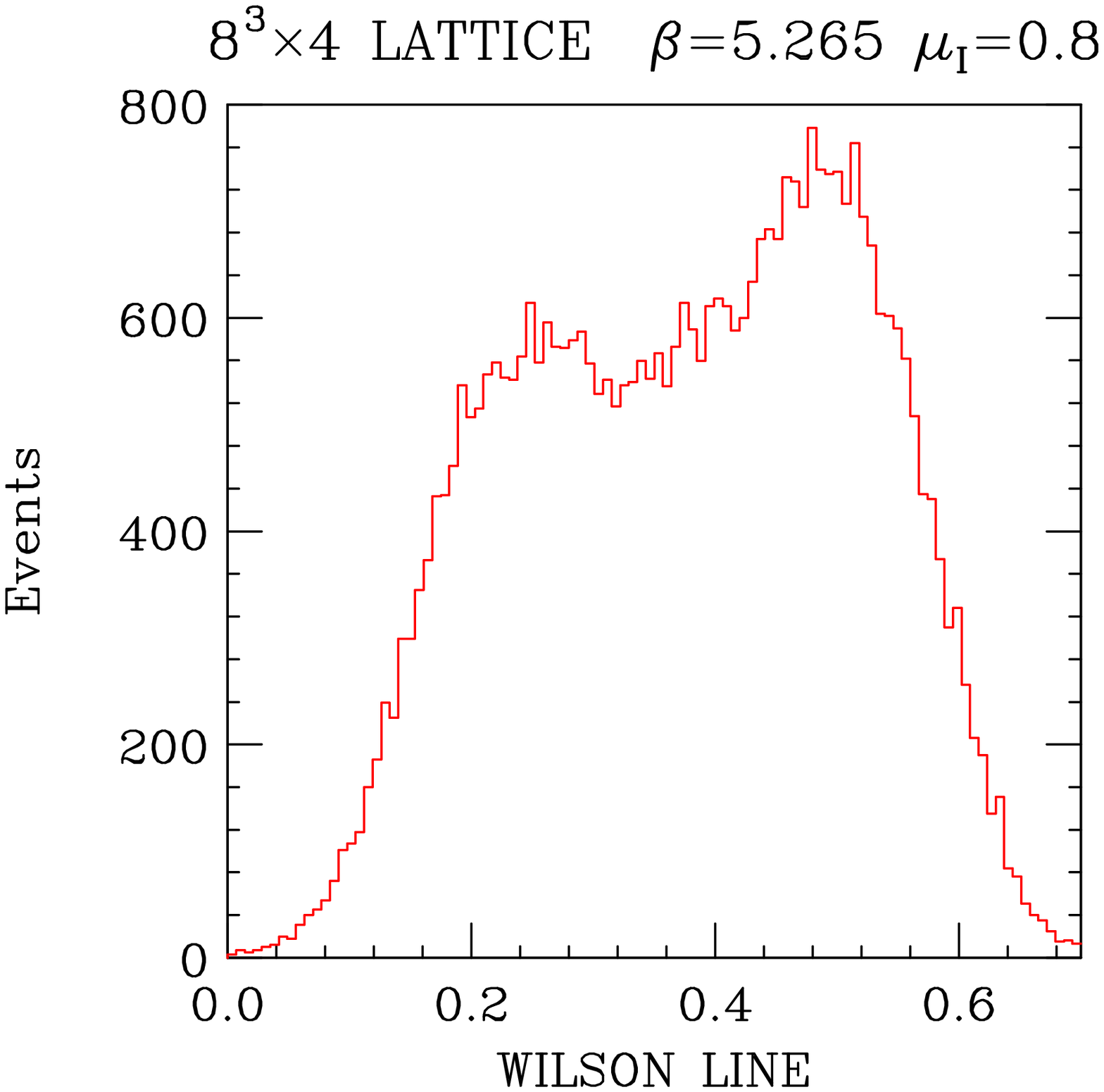}
\vspace{0.25in}
\epsfxsize=3.5in
\vspace{0.25in}
\epsffile{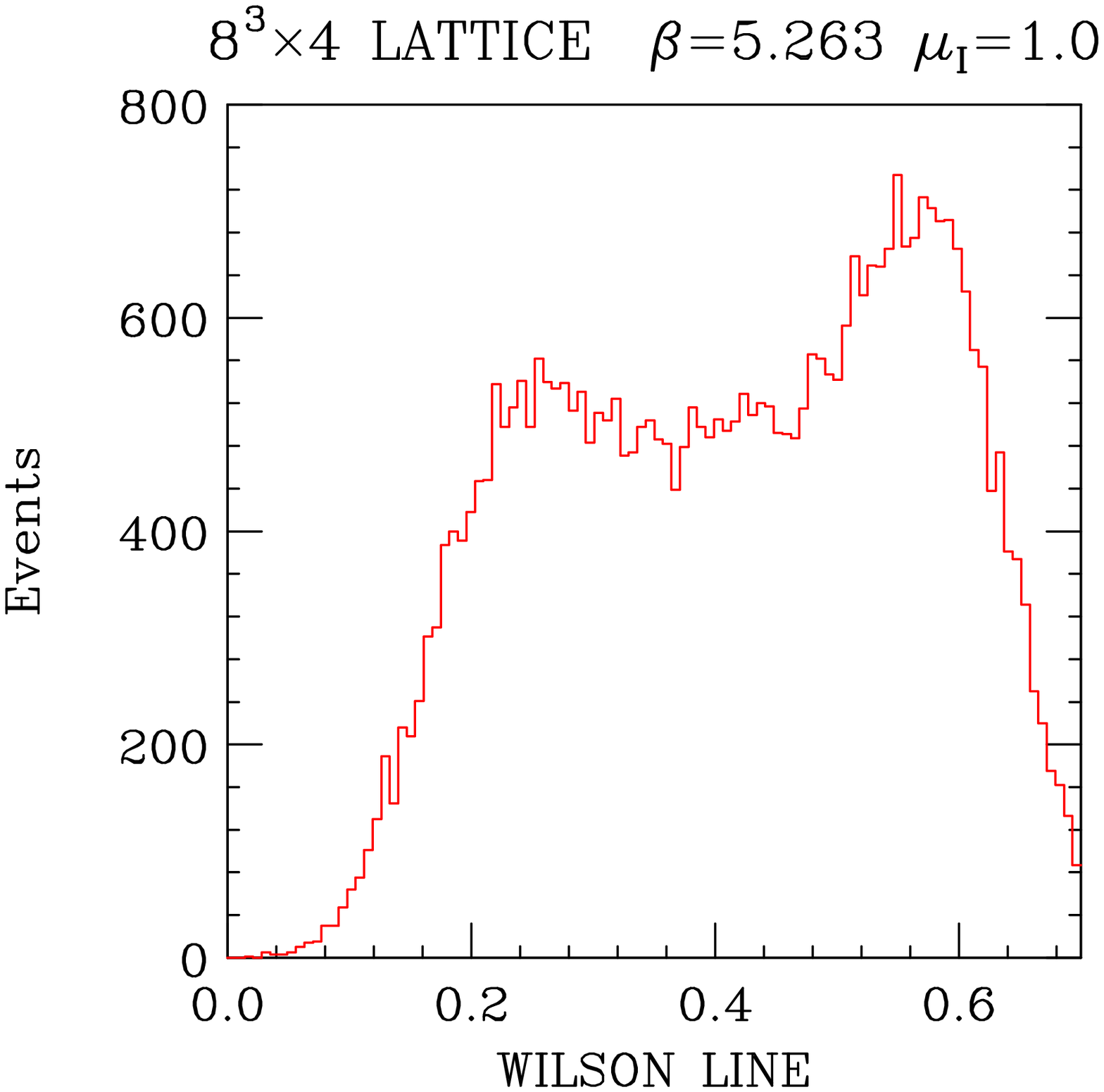}
\caption{Wilson Line histograms for large $\mu_I$ on an $8^3 \times 4$
lattice with $m=0.05$ and $\lambda=0.005$. a) $\mu_I=0.6$, $\beta=5.27$;
b) $\mu_I=0.8$, $\beta=5.265$, $dt=0.1$; c) $\mu_I=1.0$, $\beta=5.263$.}
\label{fig:hist_hi}                                                       
\end{figure}

We have performed simulations on an $8^3 \times 4$ lattice at quark mass 
$m=0.05$ and $\lambda=0.005$, at $\mu_I=0.6$, $\mu_I=0.8$ and $\mu_I=1.0$.
The $\mu_I=0.8$ simulations were repeated on a $16^3 \times 4$ lattice.  
In figure~\ref{fig:wilson_hi}, we show the behaviour of the Wilson Lines as
functions of $\beta$ for the 3 $\mu_I$ values from our simulations on an
$8^3 \times 4$ lattice.
At $\mu_I=1.0$ we have only performed simulations very close to the transition.
We obtained the high statistics needed to reveal the true nature of this
transition at those $\beta$ values closest to the transition for each $\mu_I$
--- for $\mu_I=0.6$ we obtained 40,000 time units at $\beta=5.27$, for 
$\mu_I=0.8$ we obtained 40,000 time units at $\beta=5.265$ using $dt=0.02$ and
a further 40,000 time units using $dt=0.01$, while at $\mu_I=1.0$ we obtained 
40,000 time units at $\beta=5.263$. We show histograms of the Wilson Line at
these $\beta$ and $\mu_I$ values in figure~\ref{fig:hist_hi}.
At $\mu_I=0.6$, the histogram shows no structure to suggest anything but a
crossover, which would then become a second-order transition as 
$\lambda \rightarrow 0$. By $\mu_I=0.8$ we begin to see clear signs of double
peak suggestive of a 2-state signal. The signs of a double peak and a 2-state
signal persist at $\mu_I=1.0$.

\begin{figure}[htb]
\epsfxsize=4in
\centerline{\epsffile{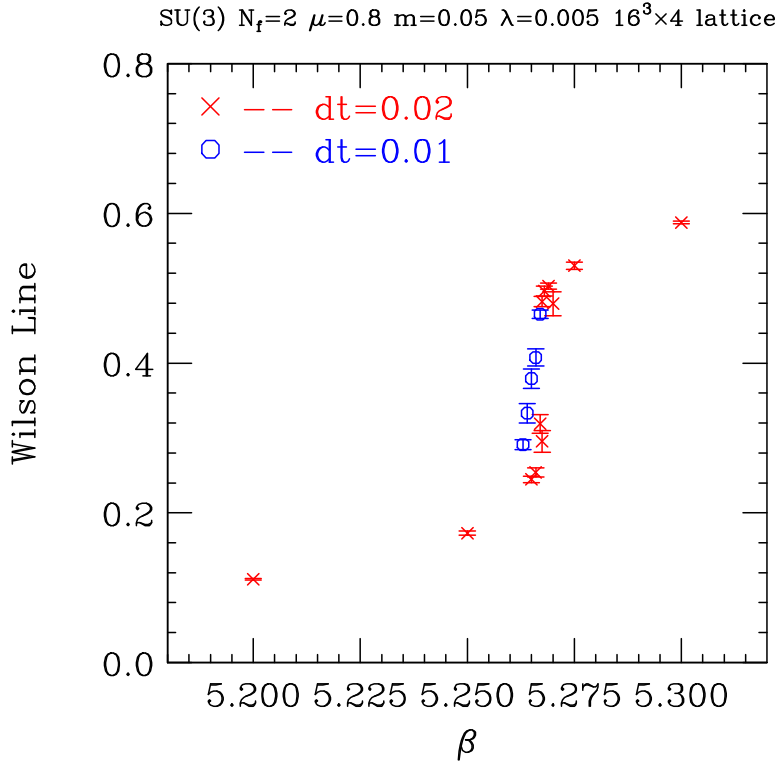}}
\vspace{0.25in}
\centerline{\epsffile{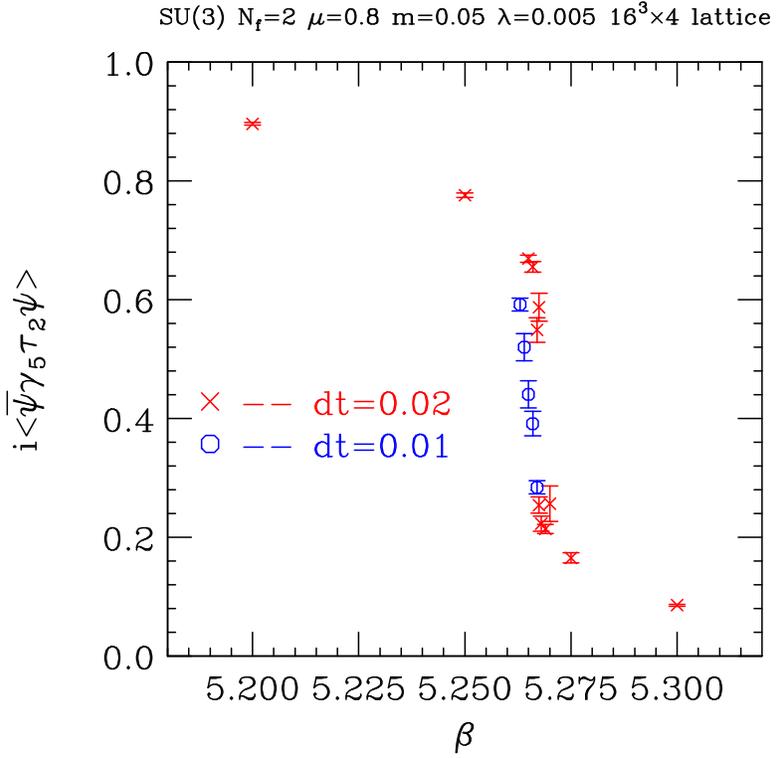}}
\caption{a) Wilson Line as a function of $\beta$ at $\mu_I=0.8$ on a 
$16^3 \times 4$ lattice. b) Charged pion condensate as a function of $\beta$
at $\mu_I=0.8$ on a $16^3 \times 4$ lattice.}\label{fig:1616164}
\end{figure}

\begin{figure}[htb]
\epsfxsize=6in
\centerline{\epsffile{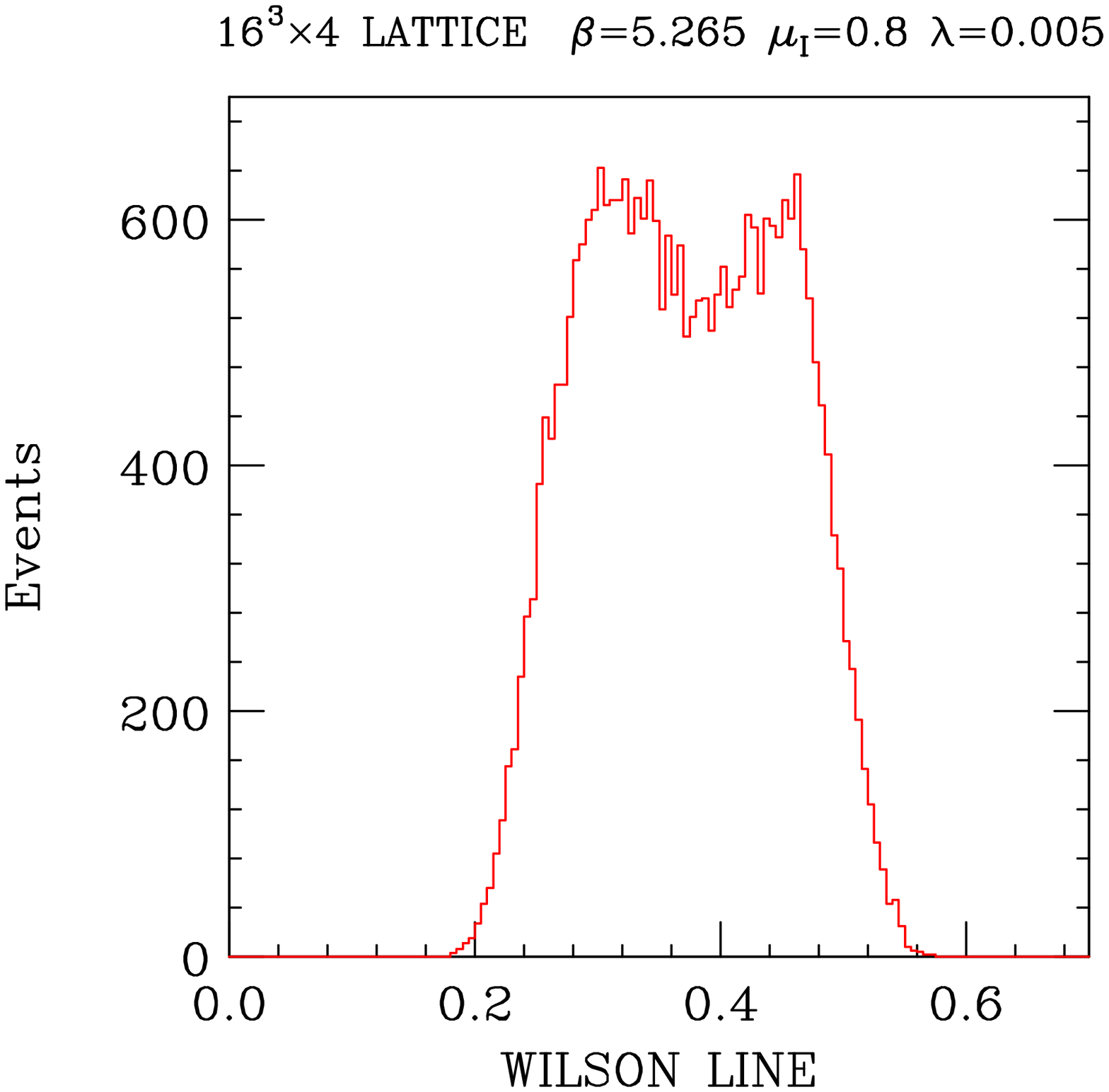}}
\caption{Histogram of Wilson line values close to the transition on a
$16^3 \times 4$ lattice at $\mu_I=0.8$ ($\beta=5.265$).}\label{fig:hist16}
\end{figure}

Since lattices as small as $8^3 \times 4$ can show signs of a 2-state signal
at a second order transition or even a crossover, we need to examine this
transition more closely. For this reason we have performed simulations on a
$16^3 \times 4$ lattice at $\mu_I=0.8$. Figure~\ref{fig:1616164} shows the
Wilson Line and the pion condensate from these simulations as functions of
$\beta$.
The reason $dt$ was decreased from $0.02$ to $0.01$ close to the transition
was that finite $dt$ effects at $dt=0.02$, both here and in our $8^3 \times 4$
runs at the same $\mu_I$, can artificially enhance the 2-state signal. $dt=0.01$
appears free from such enhancements. For our $dt=0.01$ runs at $\beta=5.263$,
$\beta=5.264$ and $\beta=5.265$ we ran for 30,000 time-units per $\beta$ to
obtain adequate statistics (for $\beta=5.266$ and $\beta=5.267$ we ran for 
20,000 time units per $\beta$). Figure~\ref{fig:hist16} shows a histogram of
the Wilson Line values from our $\beta=5.265$ runs. Although there is some
double peak structure, the peaks are considerably closer together than they
were for the $8^3 \times 4$ lattice, a sign that the double peak structure is
a finite volume artifact and not the sign of a true 2-state signal indicating
a first order transition.

\begin{figure}[htb]                                                          
\epsfxsize=6in                                                               
\centerline{\epsffile{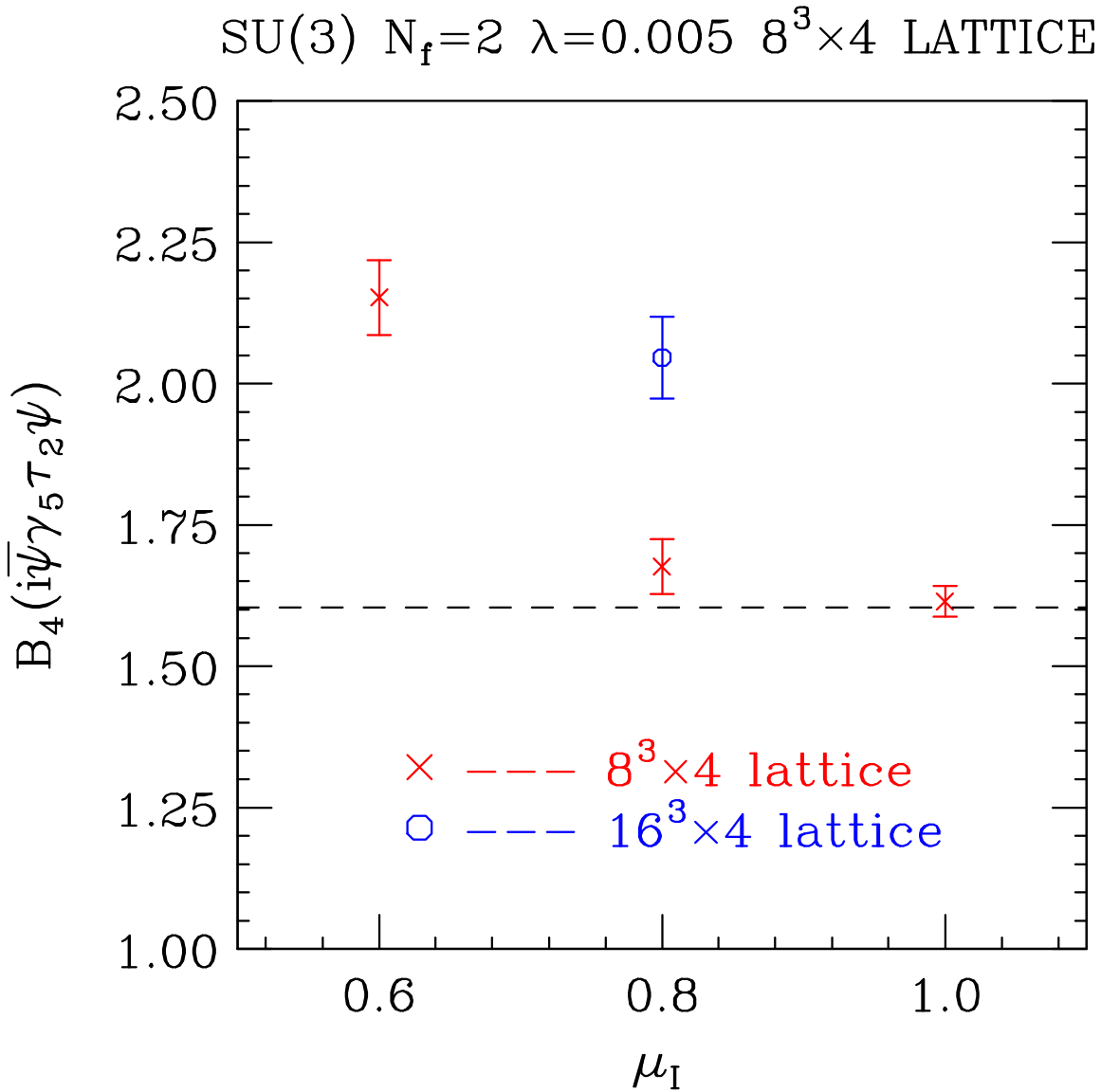}}
\caption{Binder cumulants of the charged pion condensate,
$i\langle\bar{\psi}\gamma_5\tau_2\psi\rangle$, as functions of $\mu_I$}
\label{fig:binder_l005}
\end{figure}

To clarify the situation we again turn to fourth order Binder cumulants. Here
the obvious choice is to look at the Binder cumulants of the pion condensate,
since this is the order parameter of this transition in the 
$\lambda \rightarrow 0$ limit. We plot $B_4$ versus $\mu_I$ in 
figure~\ref{fig:binder_l005} obtained using Ferrenberg-Swendsen reweighting
to obtain $B_4$ at that $\beta$ which minimizes $B_4$ for that particular
value of $\mu_I$. 
The $8^3 \times 4$ points suggest that the Binder cumulant crosses the Ising
value somewhere above $\mu_I=0.8$ and probably close to $\mu_I=1.0$. If so,
this would indicate that there is a critical end point with Ising critical
exponents at $\mu_I=\mu_c$ with $\mu_c \sim 1.0$. For $\mu_I > \mu_c$ the
transition would become first order. The $16^3 \times 4$ Binder cumulant at
$\mu_I=0.8$ is large enough to indicate that $\mu_c$ is indeed greater than
$0.8$. We would expect that if the transition is first order for
$\lambda=0.005$ it will also be first order for $\lambda=0$. Hence there will
be a tricritical point for $\mu_I=\mu_t$ where $\mu_t < \mu_c$. Since
$\lambda=0.005$ is rather small, we expect that $\mu_t \approx \mu_c$.

\section{Conclusions}

We have simulated lattice QCD with 2 flavours of staggered quarks (`half' a 
staggered quark field) with a chemical potential $\mu_I$ for isospin ($I_3$),
in the neighbourhood of the finite temperature transition. For $\mu_I < m_\pi$
we have determined the $\mu_I$ dependence of $\beta_c$, the transition 
$\beta=6/g^2$, for each of 3 quark masses. We have noted that the fluctuations
of the phase of the fermion determinant on an $8^3 \times 4$ lattice are well
enough behaved for small $\mu_I$ that there should be a range of $\mu_I$ for
which the dependence of $\beta_c$, and hence temperature, on $\mu_I$ and on the
quark-number chemical potential $\mu$ should be identical for $\mu_I=2\mu$,
as was observed previously by the Bielefeld-Swansea group for the $p$-$4$ 
action \cite{Allton:2002zi}.

What we find is that $\beta_c$ falls slowly with increasing $\mu_I$. This
falloff is approximately linear in $\mu_I^2$ over the whole $\mu_I < m_\pi$
region. The value of $\beta_c$ at $\mu_I=0$ increases with mass and the
falloff with increasing $\mu_I^2$ becomes less steep as the mass is increased.
This dependence on mass is small. We have taken the results of de Forcrand and
Philipsen \cite{deForcrand:2002ci} and converted them from a $\mu$ dependence
to a $\mu_I$ dependence. Since these were calculated at a smaller mass
$m=0.025$ than ours, we cannot compare them directly. We fit the mass
dependence of our `data' and theirs (a total of 4 different masses) to that
expected from critical scaling with both the continuum $O(4)$, and lattice
$O(2)$ universality classes, as was done at zero chemical potential in
\cite{Karsch:1994hm,Karsch:2000kv}. Both these fits prove acceptable. This is
a direct confirmation that the $\mu_I$ and $\mu$ dependence of the transition
temperature are the same, at low chemical potentials.

Since the phase of the fermion determinant is an extensive quantity, the
fluctuations in this phase increase with volume. This would suggest that the
relation between finite $\mu$ and finite $\mu_I$ transitions would fail as the
spatial volume is increased (the temporal extent is fixed at $1/T$), which is
disturbing since the infinite volume limit is of the most physical importance.
We suggest that the relevant phase to consider in establishing this 
relationship is not that on an arbitrarily large lattice, but is rather the
phase on a lattice whose spatial size is of the order of the correlation
length. Then we could limit our considerations of phase fluctuations to a
modest lattice size, unless we were very close to a critical point. 

It is worthwhile quantifying what we mean when we say that the dependence on
$\mu_I$ is slow, and what this means for the physical quantity, temperature.
If we assume that the transition temperature at $\mu_I=\mu=0$ is 
$T_c \approx 173$~MeV, then for our $m=0.05$ simulations, 2-loop running of the
coupling would imply that by $\mu_I=0.55$ or in physical units 
$\mu_I \approx 362$~MeV ($\mu \approx 181$~MeV), the transition temperature
will have fallen to $T_c \approx 164$~MeV. The relevance of this is even more
clear when one considers that this latter $\mu$ value is an order of magnitude
larger than those chemical potentials believed accessible by RHIC.

The smoothness of the transitions for $\mu_I < m_\pi$ for all 3 masses, and the
absence of any sign of a 2-state signal strongly suggests that there is no
critical endpoint, beyond which the transition would become first order, in this
domain. Analysis of the 4-th order Binder cumulant for each of the transitions,
yields values $\gtrsim 2$. Since this quantity should pass through the 3-d Ising
value $1.604(1)$ at a critical endpoint and lie below this value in the first
order region, this validates our assumption that the finite temperature
transition from hadronic matter to a quark-gluon plasma remains a crossover
throughout this region, and suggests that there is no critical endpoint for
the corresponding range of $\mu$.

We have also studied $\mu_I > m_\pi$, where the finite temperature transition
for symmetry breaking parameter $\lambda=0$ is a true phase transition from
a pion condensed superfluid to a quark-gluon plasma. Here we have performed
simulations with a small $\lambda$ ($0.1 m$), where the second order transition
for $\mu_I$ just above $m_\pi$, softens to a crossover. We see evidence that
for $\mu_I$ sufficiently large ($\sim 1$), there is a critical endpoint, where 
the 4-th order Binder cumulant passes through its 3-d Ising value, and beyond
which it is first order. This observation needs to be confirmed on larger 
lattices, where the passage through the critical endpoint is expected be 
considerably more rapid. Unfortunately, in this regime, we cannot argue that
a critical endpoint in $\mu_I$ is in any way related to a corresponding
critical endpoint in $\mu$.

We have recently begun to extend this work to the 3-flavour case. Not only is
this more physical, but one can argue that it is possible to tune the critical
endpoint to be as close to $\mu_I=0$ as one desires, by careful choice of the
quark mass $m$. In particular we can choose the critical value $\mu_I=\mu_c$
to obey $\mu_c < m_\pi$, and lie in the domain where the $\mu$ and $\mu_I$
transitions are related. Studies of the $3$ and $2+1$ flavour transitions by
various methods have located such a critical endpoint, but their predictions
of its location are not in agreement \cite{Fodor:2004nz,Karsch:2003va,
deForcrand:2003hx}.
Hence we have a chance to clarify the
situation by a more direct approach.

It has been pointed out by de Forcrand, Kim and Takaishi
\cite{deForcrand:2002pa}, that simulating with finite $\mu_I$ provides a
better ensemble for reweighting methods for finite $\mu$, than simulations
with zero chemical potential. Combined with our observations, this suggests
that such reweighting would be optimal close to the finite temperature
transition for small $\mu$, and could be expected to give good predictions for
observables in this domain. This should enable us to determine the
equation-of-state in this low-$\mu$ domain. Of course, such reweighting
requires calculating the phase of the fermion determinant. Doing this
precisely would be prohibitively expensive for all but the smallest lattices.
New methods for approximating the fermion determinant show promise for making
these reweighting methods practical \cite{Liu:2002qr}.

\subsection*{Acknowledgements}
DKS is supported under US Department of Energy, Division of High Energy
Physics, contract W-31-109-ENG-38. JBK is supported in part by a
National Science Foundation grant NSF PHY03-04252. The simulations described
in this paper were performed on the IBM SP, Seaborg, at NERSC, the Jazz Linux
PC cluster at the LCRC, Argonne National Laboratory, and Linux PCs belonging
the HEP Division at Argonne National Laboratory. One of us (DKS) would like
to thank Philippe de Forcrand for helpful discussions about their work, its
relationship to that of the Bielefeld-Swansea group and on the utility of
Binder cumulant analyses of transitions. He would also like to thank Simon
Hands for discussions of the work of the Bielefeld-Swansea group.


\begin{thebibliography}{999}

\bibitem{Fodor:2001au}
Z.~Fodor and S.~D.~Katz,
Phys.\ Lett.\ B {\bf 534}, 87 (2002)
[arXiv:hep-lat/0104001].

\bibitem{Fodor:2001pe}
Z.~Fodor and S.~D.~Katz,
JHEP {\bf 0203}, 014 (2002)
[arXiv:hep-lat/0106002].

\bibitem{Fodor:2004nz}
Z.~Fodor and S.~D.~Katz,
JHEP {\bf 0404}, 050 (2004)
[arXiv:hep-lat/0402006].

\bibitem{Allton:2002zi}
C.~R.~Allton {\it et al.},
Phys.\ Rev.\ D {\bf 66}, 074507 (2002)
[arXiv:hep-lat/0204010].

\bibitem{Karsch:2003va}
F.~Karsch, C.~R.~Allton, S.~Ejiri, S.~J.~Hands, O.~Kaczmarek, E.~Laermann and C.~Schmidt,
Nucl.\ Phys.\ Proc.\ Suppl.\  {\bf 129}, 614 (2004)
[arXiv:hep-lat/0309116].

\bibitem{deForcrand:2002ci}
P.~de Forcrand and O.~Philipsen,
Nucl.\ Phys.\ B {\bf 642}, 290 (2002)
[arXiv:hep-lat/0205016].

\bibitem{deForcrand:2003hx}
P.~de Forcrand and O.~Philipsen,
Nucl.\ Phys.\ B {\bf 673}, 170 (2003)
[arXiv:hep-lat/0307020].

\bibitem{D'Elia:2002gd}
M.~D'Elia and M.~P.~Lombardo,
Phys.\ Rev.\ D {\bf 67}, 014505 (2003)
[arXiv:hep-lat/0209146].

\bibitem{Gavai:2003mf}
R.~V.~Gavai and S.~Gupta,
Phys.\ Rev.\ D {\bf 68}, 034506 (2003)
[arXiv:hep-lat/0303013].

\bibitem{Gavai:2003nn}
R.~Gavai, S.~Gupta and R.~Roy,
Prog, Theor. Phys. Suppl. {\bf 153}, 270 (2004)
[arXiv:nucl-th/0312010].

\bibitem{Azcoiti}
V.~Azcoiti, Talk presented at Lattice2004, Fermilab (2004)

\bibitem{Son:2000xc}
D.~T.~Son and M.~A.~Stephanov,
Phys.\ Rev.\ Lett.\  {\bf 86}, 592 (2001)
[arXiv:hep-ph/0005225].

\bibitem{Son:2000by}
D.~T.~Son and M.~A.~Stephanov,
Phys.\ Atom.\ Nucl.\  {\bf 64}, 834 (2001)
[Yad.\ Fiz.\  {\bf 64}, 899 (2001)]
[arXiv:hep-ph/0011365].

\bibitem{Kogut:2002zg}
J.~B.~Kogut and D.~K.~Sinclair,
Phys.\ Rev.\ D {\bf 66}, 034505 (2002)
[arXiv:hep-lat/0202028].

\bibitem{Kogut:2002se}
J.~B.~Kogut and D.~K.~Sinclair,
Nucl.\ Phys.\ Proc.\ Suppl.\  {\bf 119}, 556 (2003)
[arXiv:hep-lat/0209054].

\bibitem{Kogut:2003cd}
J.~B.~Kogut and D.~K.~Sinclair,
Nucl.\ Phys.\ Proc.\ Suppl.\  {\bf 129}, 542 (2004)
[arXiv:hep-lat/0309042].

\bibitem{Sinclair:2003rm}
D.~K.~Sinclair, J.~B.~Kogut and D.~Toublan,
Prog. Theor. Phys. Suppl. {\bf 153}, 40 (2004)
[arXiv:hep-lat/0311019].

\bibitem{binder} 
K.~Binder, Z.~Phys. B {\bf 43}, 119 (1981).

\bibitem{Karsch:2001nf}
F.~Karsch, E.~Laermann and C.~Schmidt,
Phys.\ Lett.\ B {\bf 520}, 41 (2001)
[arXiv:hep-lat/0107020].

\bibitem{lipowski}
A.~Lipowski and M.~Droz, Phys. Rev. E {\bf 66}, 016118 (2002).

\bibitem{Ferrenberg:yz}
A.~M.~Ferrenberg and R.~H.~Swendsen,
Phys.\ Rev.\ Lett.\  {\bf 61}, 2635 (1988).

\bibitem{Hatta:2003ga}
Y.~Hatta and K.~Fukushima,
arXiv:hep-ph/0307068.

\bibitem{Fukushima:2002mp}
K.~Fukushima,
Phys.\ Rev.\ C {\bf 67} (2003) 025203
[arXiv:hep-ph/0209270].

\bibitem{Mocsy:2003tr}
A.~Mocsy, F.~Sannino and K.~Tuominen,
Phys.\ Rev.\ Lett.\  {\bf 91} (2003) 092004
[arXiv:hep-ph/0301229].

\bibitem{Mocsy:2003qw}
A.~Mocsy, F.~Sannino and K.~Tuominen,
arXiv:hep-ph/0308135.

\bibitem{Takaishi:2004qa}
T.~Takaishi,
Prog. Theor. Phys. Suppl. {\bf 153}, 277 (2004)
[arXiv:hep-lat/0405010].

\bibitem{Karsch:1994hm}
F.~Karsch and E.~Laermann,
Phys.\ Rev.\ D {\bf 50}, 6954 (1994)
[arXiv:hep-lat/9406008].

\bibitem{Karsch:2000kv}
F.~Karsch, E.~Laermann and A.~Peikert,
Nucl.\ Phys.\ B {\bf 605}, 579 (2001)
[arXiv:hep-lat/0012023].

\bibitem{deForcrand:2002pa}
P.~de Forcrand, S.~Kim and T.~Takaishi,
Nucl.\ Phys.\ Proc.\ Suppl.\  {\bf 119}, 541 (2003)
[arXiv:hep-lat/0209126].

\bibitem{Liu:2002qr}
K.~F.~Liu,
Int.\ J.\ Mod.\ Phys.\ B {\bf 16}, 2017 (2002)
[arXiv:hep-lat/0202026].

\end{thebibliography}
\end{document}